\documentclass[aps,prl,reprint,groupedaddress]{revtex4-2}
\usepackage{graphicx} 
\usepackage{amsmath} 
\usepackage{amssymb} 
\usepackage{hyperref} 
\usepackage{xcolor} 
\usepackage{float}
\usepackage{comment}
\usepackage{soul}

\begin{document}



\title{ Simulation Studies of Resonant Excitation of Electron Bernstein Waves in Capacitive Discharges}

\author{Deepak Gautam$^1$}
\author{Sarveshwar Sharma$^{2,3,}$}
\email{sarvesh@ipr.res.in, sarvsarvesh@gmail.com}
\author{Igor Kaganovich$^4$}
\author{Bhooshan Paradkar$^1$}


\affiliation{$^1$ School of Physical Sciences, UM-DAE Centre for Excellence in Basic Sciences, University of Mumbai, Mumbai 400098, India}
\affiliation{$^2$Institute for Plasma Research, Bhat, Gandhinagar, Gujarat 382428, India}
\affiliation{$^3$Homi Bhabha National Institute, Training School Complex, Anushaktinagar, Mumbai 400094, India}
\affiliation{$^4$Princeton Plasma Physics Laboratory, Princeton, New Jersey 08543, USA}

\date{\today}

\begin{abstract}
The behavior of capacitive coupled plasma (CCP) discharges is investigated in a mildly magnetized regime, defined by the condition 
$\mathrm{1 \leq f_{ce}/f_{rf} < 2}$, where $\mathrm{f_{ce}}$ and $\mathrm{f_{rf}}$ are the cyclotron and radio-frequencies (RF), respectively. This regime exhibits complex and distinctive plasma dynamics due to the interplay between RF fields and the externally applied magnetic field. Two prominent phenomena are observed in this regime. First, the plasma density profile becomes asymmetric across the discharge, deviating from the typical symmetric distribution seen in unmagnetized CCPs. Second, electron Bernstein waves (EBWs), high-frequency electrostatic waves, are excited and propagate within the bulk plasma, particularly along steep electron density gradients. As the strength of the magnetic field increases within this regime, the CCP discharge undergoes a transition from a symmetric configuration to an asymmetric one, and then returns to a symmetric profile at higher field strengths. Notably, the excitation and propagation of EBWs are strongly correlated with the presence of discharge asymmetry and localized density gradients. These waves play a significant role in energy transport and electron heating under mildly magnetized conditions.
To gain deeper insight into the underlying physics, detailed numerical simulations are carried out using the particle-in-cell Monte Carlo collision (PIC-MCC) technique. These simulations capture the kinetic behavior of electrons and ions, including the collisionless effects and sheath dynamics essential to understanding the excitation of EBWs and the evolution of discharge symmetry. The study thus sheds light on the role of weak magnetic fields in shaping plasma behavior and highlights the importance of wave-particle interactions in magnetized CCPs.

\end{abstract}

\maketitle

\section{1. Introduction}
\begin{figure}
    \centering
    \includegraphics[scale=0.30]{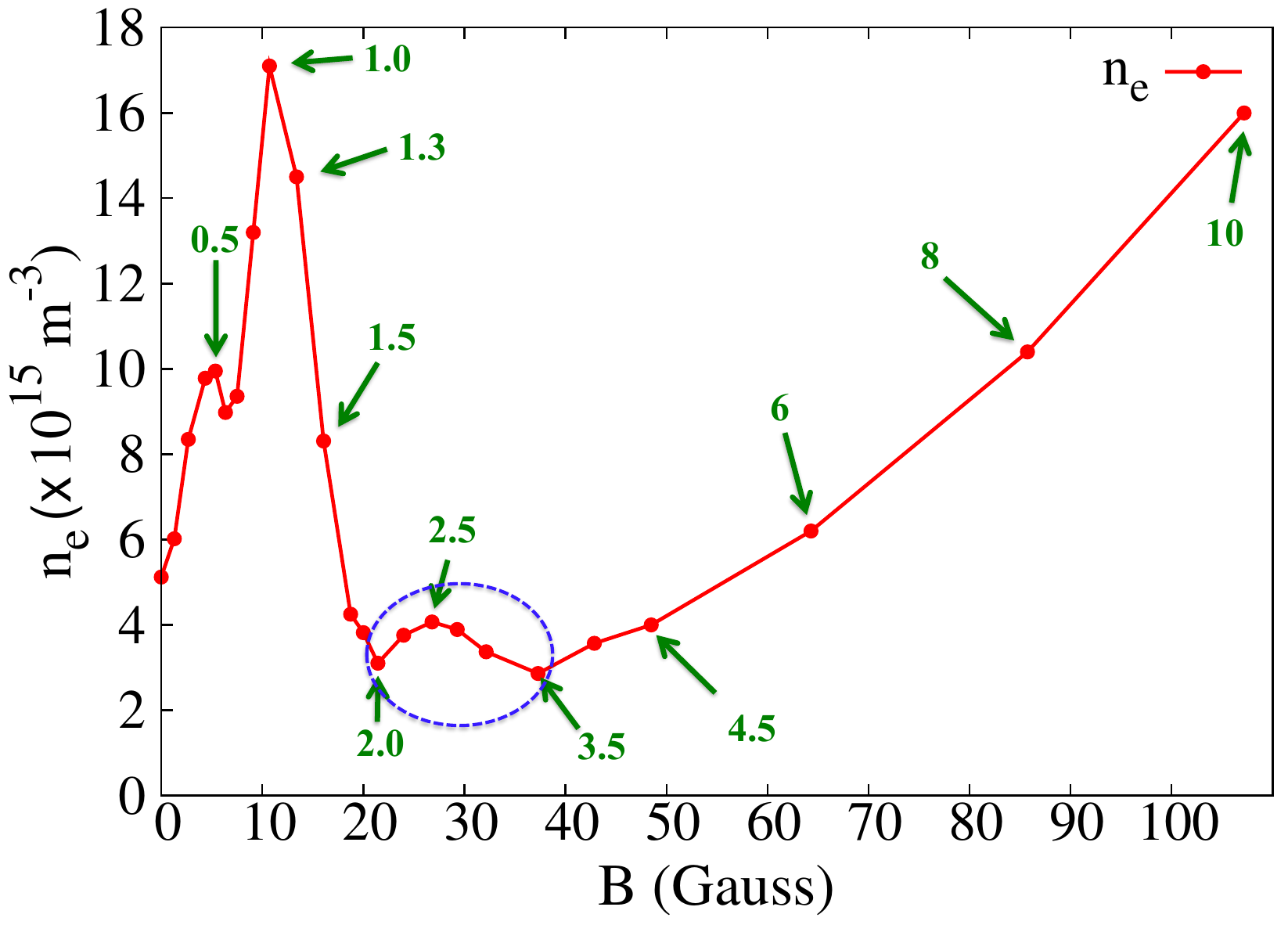} 
    \caption{Peak plasma density with varying magnetic field strengths as reported by Patil et al.\cite{patil2022electron}. The values of $\mathrm{r}$, corresponding to the given magnetic field strength, are shown by green arrows. Present studies focus on the physics of transition region ($\mathrm{2.0 \leq r \leq 3.5}$), which is circle by a blue dotted line.
    }
    \label{3Regions}
\end{figure}

Low-pressure radio-frequency (RF) capacitive coupled plasma (CCP) discharges are the foundation for a wide range of advanced technologies with significant societal and industrial applications\cite{article,chabert2011physics,book}. These discharges are extensively used in fields such as semiconductor manufacturing, surface coating, biomedical engineering, and environmental applications\cite{coburn1979ion,gottscho1992microscopic,hopwood1992review,kong2009plasma}.
Especially in the context of modern semiconductor fabrication, plasma-based processes such as ion-assisted etching \cite{coburn1979ion,gottscho1992microscopic} and ion implantation \cite{hopwood1992review,kong2009plasma,madou2011manufacturing} are particularly vital.\\
In semiconductor processing, achieving nanoscale precision and high-aspect-ratio structures requires highly controlled plasma etching techniques. Deep reactive-ion etching (DRIE) is an advanced anisotropic dry etching method capable of producing aspect ratios up to 50:1. It enables nearly vertical trench profiles with sidewall angles close to 90°, making it essential for next-generation microelectronic and MEMS devices \cite{laermer1996method,hoenk1992growth}.\\
Plasma tools also support integration depths of just a few nanometers, allowing fabrication at dimensions that meet the stringent requirements of modern device architectures \cite{hoenk1992growth}. However, a key challenge in these processes is the independent control of ion energy and ion flux at the substrate surface. This control is crucial for tailoring the etch profiles, minimizing substrate damage, and ensuring uniformity between large wafers\cite{lieberman1994principles,chabert2011physics,makabe2006plasma,heil2008possibility,turner2006collisionless}.\\
In single-frequency capacitively coupled plasmas (SF-CCPs), the ion energy and ion flux on the substrate are inherently coupled \cite{popov1985power,lieberman1988analytical,kaganovich2006revisiting,kawamura2006stochastic,sharma2013simulation,sharma2013simulation2}. To overcome this, various control techniques have been developed. A widely used approach is dual-frequency CCP, where a high-frequency source sustains plasma and controls ionization, while a low-frequency source regulates sheath voltage and ion energy, allowing decoupling of ion energy and flux \cite{robiche2003analytical,boyle2004independent,turner2006collisionless,sharma2013critical,sharma2014investigation,sharma2013investigation,turner2006collisionless,gans2006frequency}. Additionally, tailored non-sinusoidal or asymmetric voltage waveforms can manipulate sheath dynamics to control ion energies without significantly affecting plasma density \cite{economou2013tailored,lafleur2015tailored,shin2011control,sharma2015collisionless,schungel2015customized,sharma2020high,sharma2021ion,sharma2024harmonic,sharma2023spatial} [35–47]. Other methods, such as asymmetric electrode designs and DC or RF biasing, are also used to tailor ion energy distributions on the substrate \cite{qin2010tailored,upadhyay2015reversal}.\\
Recent research has focused on very-high-frequency (VHF, 30–300 MHz) CCP discharges. VHF operation improves plasma performance by producing higher plasma densities at a given power due to increased plasma current, while also reducing DC self-bias, leading to higher etching rates with less substrate damage compared to low-frequency discharges \cite{rauf2009power,wilczek2015effect,sharma2016effect}. Additionally, VHF CCPs exhibit phenomena such as electric field transients in the bulk and sheath regions, and higher harmonics in voltage and current waveforms \cite{upadhyay2013effect,sharma2019electric,miller2006spatial,wilczek2018disparity,sharma2019influence,sharma2020electric,simha2023kinetic}. These effects significantly influence electron heating mechanisms, plasma uniformity, and overall process performance.\\
Furthermore, the application of external magnetic fields has been extensively explored to enhance the control over ion dynamics, particularly within the
framework of magnetically enhanced reactive ion etching (MERIE)\cite{muller1989magnetically,hutchinson1995effects,kushner2003modeling,vasenkov2004modeling}.
You et al.\cite{you2011role} experimentally investigated the influence of a static transverse magnetic field on asymmetric single-frequency CCP argon discharges driven at 13.56 MHz. Their study, conducted in the low to intermediate pressure regime, revealed the presence of $\mathrm{E \times B}$ drift, where the electric field (E) perpendicular to the electrodes and the magnetic field (B) parallel to the electrodes interact to modify electron trajectories. Yang et al. \cite{yang2018magnetical} showed that a magnetic field with an asymmetric spatial gradient can induce discharge asymmetry, enabling independent control of ion energy and flux. Sharma et al. \cite{sharma2018spatial}, using the electrostatic direct implicit particle-in-cell (EDIPIC) code \cite{sydorenko2006particle}, further demonstrated that a uniform transverse magnetic field can enhance ion flux while effectively controlling ion energy. These results highlight the potential of external magnetic fields for tailoring plasma characteristics in SF-CCP systems.\\
Strong magnetic fields (tens to hundreds of Gauss) are known to introduce plasma non-uniformities due to $\mathrm{E \times B}$ drift, adversely affecting wafer-scale uniformity \cite{barnat2008rf,fan2013study}. For example, Barnat et al. \cite{barnat2008rf} observed such non-uniformities in argon CCPs at 13.56 MHz. Similarly, Fan et al. \cite{fan2013study}, using 2D-3V PIC simulations at 20 mTorr and 13.56 MHz with 200 V, showed that increasing the magnetic field (0–50 G) enhances plasma density, but significant non-uniformities arise beyond ~20 G.\\
In low-pressure CCP discharges, the introduction of an external magnetic field can excite waves that contribute to bulk plasma heating via wave–particle interactions. Among these, electron Bernstein waves (EBWs) are particularly important, as they can propagate in magnetized, overdense plasmas where electromagnetic waves (O and X-modes) become evanescent when 
$\mathrm{\omega_{pe}}$ exceeds the wave frequency. Due to their electrostatic nature and unique dispersion properties, EBWs can penetrate overdense regions and enable efficient energy transfer where conventional wave heating is ineffective \cite{preinhaelter1973penetration, sugai1981mode}. EBWs are of particular interest in CCPs due to their collisionless heating capacity, especially under low-pressure conditions (typically $<$ 10 mTorr) where traditional collisional (ohmic) heating mechanisms are less effective \cite{article}. These waves are excited at harmonics of the electron cyclotron frequency $\mathrm{\omega_{ce} = eB/m_e}$, and are primarily driven by the perpendicular thermal motion of electrons with respect to the magnetic field. Although the possibilities of heating plasma with EBWs have been extensively explored in fusion devices, these waves remain largely unexplored in the magnetized CCP discharges.\\
More recently, a novel regime has been identified in which low-pressure SF-CCP discharges operating at VHF show enhanced performance under the influence of relatively weak transverse magnetic fields\cite{patil2020enhanced,patil2022electron,zhang2021resonant,
sharma2022investigating}. Patil et al.\cite{patil2020enhanced,patil2022electron},using PIC simulations, were the first to demonstrate considerable performance enhancement by applying a transverse magnetic field of nearly 10 G to a $\mathrm{5 \  mTorr}$ discharge driven at 60 MHz. This improvement was attributed to a resonance condition in which the electron cyclotron frequency $\mathrm{f_{ce}}$ becomes equal to half the applied RF frequency $\mathrm{f_{rf}}$. This phenomenon, referred to as Electron Bounce Cyclotron Resonance (EBCR), results in more efficient electron heating and improved plasma sustaination. Subsequent studies by Zhang et al.\cite{zhang2021resonant} confirmed these results across a broader frequency range (13.56–60 MHz) using both PIC simulations and experimental validation. 
Furthermore, Sharma et al.\cite{sharma2022investigating} have provided a detailed investigation of the mechanisms behind EBCR and its broader implications on plasma properties. These techniques have significantly advanced the capability of RF-CCP discharges to meet the stringent demands of modern plasma processing.\\
A typical trend in peak plasma density with varying magnetic field, as reported by Patil et al.\cite{patil2022electron}, is shown in fig.\ref{3Regions}. In this figure, the strength of external magnetic field is quantified through a non-dimensional number $r$, defined as:
\begin{equation}
\label{ratio}
    r = \dfrac{\mathrm{2 f_{ce}}}{\mathrm{f_{rf}}}=\dfrac{\mathrm{2 \omega_{ce}}}{\mathrm{\omega_{rf}}}.
\end{equation}
Here, $\mathrm{f_{ce}}$ and $\mathrm{f_{rf}}$ represent the electron cyclotron frequency and the applied RF frequency, respectively. The values of $r$, corresponding to a given magnetic field strength, are shown by green arrows. Thus, density peak associated with EBCR appears at $r = 1$ i.e. when $\mathrm{f_{ce}}$ becomes equal to half of the applied RF frequency $\mathrm{f_{rf}}$. However, as the strength of the externally applied magnetic field increases beyond the resonance point, the plasma density initially decreases before exhibiting a secondary rise near $\mathrm{r = 2.5}$. Notably, this second density peak is significantly lower in magnitude compared to that observed at resonance. Furthermore, increasing the magnetic field beyond a critical value ($\mathrm{r > 3.5}$) the plasma density continues to increase. This is typically attributed to the strong magnetization of plasma\cite{muller1989magnetically,hutchinson2002effects,fan2013study}. Thus, there appears to be a regime in which the mildly magnetized discharge transitions to the strongly magnetized discharge. The secondary density peak, mentioned earlier, seems to be occurring in this transition regime. In the present work, we delve deeper into the underlying physics responsible for the emergence of this secondary density peak. Through detailed analysis and simulations, we aim to build upon the findings of Patil et al. \cite{patil2020enhanced,patil2022electron} and provide further insight into plasma behavior beyond the EBCR condition.\\
As we explore the regime beyond EBCR, in the range $\mathrm{2 < r < 3.5}$, the fundamental harmonic of the electron cyclotron frequency $\mathrm{f_{ce}}$ can be efficiently driven by the RF frequency $\mathrm{f_{rf}}$ at $\mathrm{r = 2}$. Since EBWs are excited at the harmonics of $\mathrm{f_{ce}}$, favorable conditions exist for the presence of these waves. We observe that in this regime the electric field penetration in the bulk plasma happens through excitation of EBWs. The present study also demonstrates the possibility of exciting EBW in low-pressure CCP discharges. Through a comprehensive spectral analysis of the transient electric field within the bulk plasma, we provide definitive evidence for the existence of EBWs. This analysis reveals distinct wave signatures that align with the theoretical dispersion characteristics of EBWs in magnetized plasma environments. The observed frequency components and their spatial distribution within the plasma bulk are consistent with the propagation of electrostatic waves in the regime. \\ 
The structure of this paper is organized as follows: Section II presents the simulation methodology, including a detailed description of the numerical
techniques and the parameters used in the study. Section III is dedicated to the results and discussion, where key findings are analyzed and interpreted in the
context of relevant physical mechanisms. Finally, Section IV provides the conclusions, summarizing the main outcomes of the work and outlining their potential implications.
\section{2. \ Simulation Technique And Parameters}
In this study, we employ the well-established and extensively validated 1D-3V electrostatic Direct Implicit Particle-in-Cell (EDIPIC) code\cite{sun2023direct,simha2023kinetic,sharma2018spatial,sydorenko2006particle,campanell2012instability,sheehan2013kinetic,
charoy20192d}. Comprehensive information and EDIPIC source code can be accessed on GitHub \cite{edipic_github}. Although EDIPIC is capable of operating in both explicit and implicit modes, all simulations presented in this work have been performed using an explicit scheme. The computational domain models a plasma bounded by two parallel plate electrodes, simulating a planar discharge configuration. The code is built on the robust PIC-MCC framework\cite{birdsall1991plasma,hockney2021computer}, which enables the self-consistent treatment of the dynamics of charged particles alongside collisional processes. Electron-neutral collisions incorporated in the simulations include elastic scattering, excitation, and ionization events. For ion-neutral interactions, both elastic and charge exchange collisions are considered. Metastable reactions, while relevant under some conditions, are deemed negligible at low pressures considered in this study and are therefore not tracked.\\
Collision cross-section data are sourced from reliable and widely accepted databases\cite{rauf1997argon,lauro2004analysis}, ensuring the physical accuracy of the modeled interactions. The code tracks the evolution of the positions and velocities of electrons and singly ionized argon ions ($\mathrm{Ar^{+}}$). Although the spatial domain is one-dimensional (1D), the inclusion of three velocity components (3V) enables the accurate resolution of complex velocity-space dynamics, including $\mathrm{E \times B}$ drifts. This feature is critical for capturing the essential physics of magnetized plasma behavior within the scope of this work. In the present simulations, the dynamics of the neutral gas is not explicitly evolved. Instead, the neutral gas is assumed to have a uniform spatial distribution across the region between the electrodes and remains unchanged throughout the simulation duration. The neutral gas temperature is kept constant at 300 K, representing room-temperature conditions.\\
Secondary electron emission from the electrode surfaces was neglected in this study. This simplification is justified because, under low-pressure conditions, the contribution of secondary electrons to the overall discharge dynamics is minimal and does not significantly influence key plasma parameters\cite{article}. Furthermore, the effects of the external driving circuit are not included in the simulation model. Instead of modeling the complex interaction between the plasma and external circuit elements, a fixed voltage is applied directly to the electrodes. This approach allows the current waveform to naturally evolve in response to the plasma conditions, simplifying the computational setup while still capturing the essential physics of the discharge behavior.\\
In the present study, argon is used as the background neutral gas at a pressure of 5 mTorr, which is a typical low-pressure plasma processing condition. The discharge is driven by an RF voltage with a frequency of 60 MHz and a peak amplitude of 100 V. This RF voltage is applied across two parallel plate electrodes, a grounded electrode (G.E.) and a powered electrode (P.E.), which are separated by a distance of 32 mm. The powered electrode is subjected to a voltage waveform that changes over time, while the grounded electrode is maintained at zero potential. The applied voltage is mathematically expressed as
    \begin{equation}
        \label{eq1}
        \mathrm{V_{rf}(t)= V_0 \ \text{Sin}(2\pi f_{rf} t + \phi)}
    \end{equation}

where $\mathrm{V_0}$ is the peak voltage (100 V) , $\mathrm{f_{rf}}$ is the RF frequency (60 MHz) , t is time and $\phi$ is the phase angle. For all simulations presented in this paper, $\mathrm{\phi}$ is set to zero. This time-dependent waveform plays a crucial role in driving the plasma dynamics, influencing processes such as sheath formation, electron heating, and ion acceleration within the discharge gap.\\
The assumption of a one-dimensional (1D) spatial domain is justified for this study, as the primary plasma behavior of interest, driven by parallel plate geometry and symmetric conditions, can be effectively captured along the axial direction between the electrodes. An external magnetic field $B$, oriented parallel to the electrode surfaces, is systematically varied from 20 ($r = 2.0$) to 75 G ($r = 3.5$)  to explore its influence on plasma behavior. The initial electron and ion temperatures are set to 2 eV and 0.026 eV (corresponding to 300 K), respectively. The neutral gas temperature is assumed to be equal to the ion temperature. \\
To ensure accurate resolution of plasma characteristics, the spatial grid cell size $\Delta \mathrm{x}$ is chosen to be one-sixteenth of the Debye length $\mathrm{\lambda_D}$, defined as $\mathrm{\lambda_D=\sqrt{\varepsilon_0 T_e/n_0e}}$, where $\varepsilon_0$ is the vacuum permittivity, $T_e$ is the electron temperature in electron volt , $\mathrm{n_0=5\times 10^{15} \ \mathrm{m}^{-3}}$ is the reference plasma density, and $e$ is the elementary charge. For these parameters, the resulting grid cell size is approximately $\Delta \mathrm{x} = 1.314 \times 10^{-5} \ \mathrm{m}$, sufficiently small to resolve the Debye length. Given the total system length of 32 mm, this corresponds to 2436 grid cells.\\
The temporal resolution is determined on the basis of the maximum expected particle velocity, taken as four times the electron thermal velocity. The time step $\Delta \mathrm{t}$ is then calculated as
\begin{equation}
    \label{eq2}
    \Delta \mathrm{t} = \dfrac{\Delta \mathrm{x}}{ \ \mathrm{v_{max}}} \approx 3.916 \times 10^{-12} \mathrm{s}
\end{equation}
which satisfies the numerical stability condition $\mathrm{\omega_{pe}} \Delta \mathrm{t} < 0.2$, where $\mathrm{\omega_{pe}}  \Delta \mathrm{t}=0.0156$, ensuring stable integration of the particle motion. Furthermore, this time step allows for adequate resolution of the electron cyclotron motion at the maximum applied magnetic field (B=75G). The corresponding electron cyclotron period $\mathrm{T_{ce}}$ is given by
\begin{equation}
    \label{eq3}
    \mathrm{T_{ce}}=\dfrac{1}{\ \mathrm{f_{ce}}}=\dfrac{2 \pi \mathrm{m_{e}}}{\mathrm{eB}} \approx 4.75\times 10^{-9} \mathrm{s} \approx 1213.2 \ \Delta \mathrm{t} 
\end{equation}
where $\mathrm{f_{ce}}$ is the electron cyclotron frequency, $\mathrm{m_e}$ is the electron mass, and $\mathrm{B}$ is the magnetic field strength. This confirms that the cyclotron motion is well-resolved even at the highest field strength.\\
The simulation is initialized with 400 super particles per cell, yielding a total particle count of approximately one million. At the boundaries, perfect absorbing conditions are applied and charged particles are removed from the system upon reaching either electrode, mimicking realistic electrode interaction without introducing artificial reflections. The simulations were run for more than 6000 RF cycles to ensure that the system reached steady-state condition.
\begin{figure*}
    \centering
    \includegraphics[width=0.9 \textwidth]{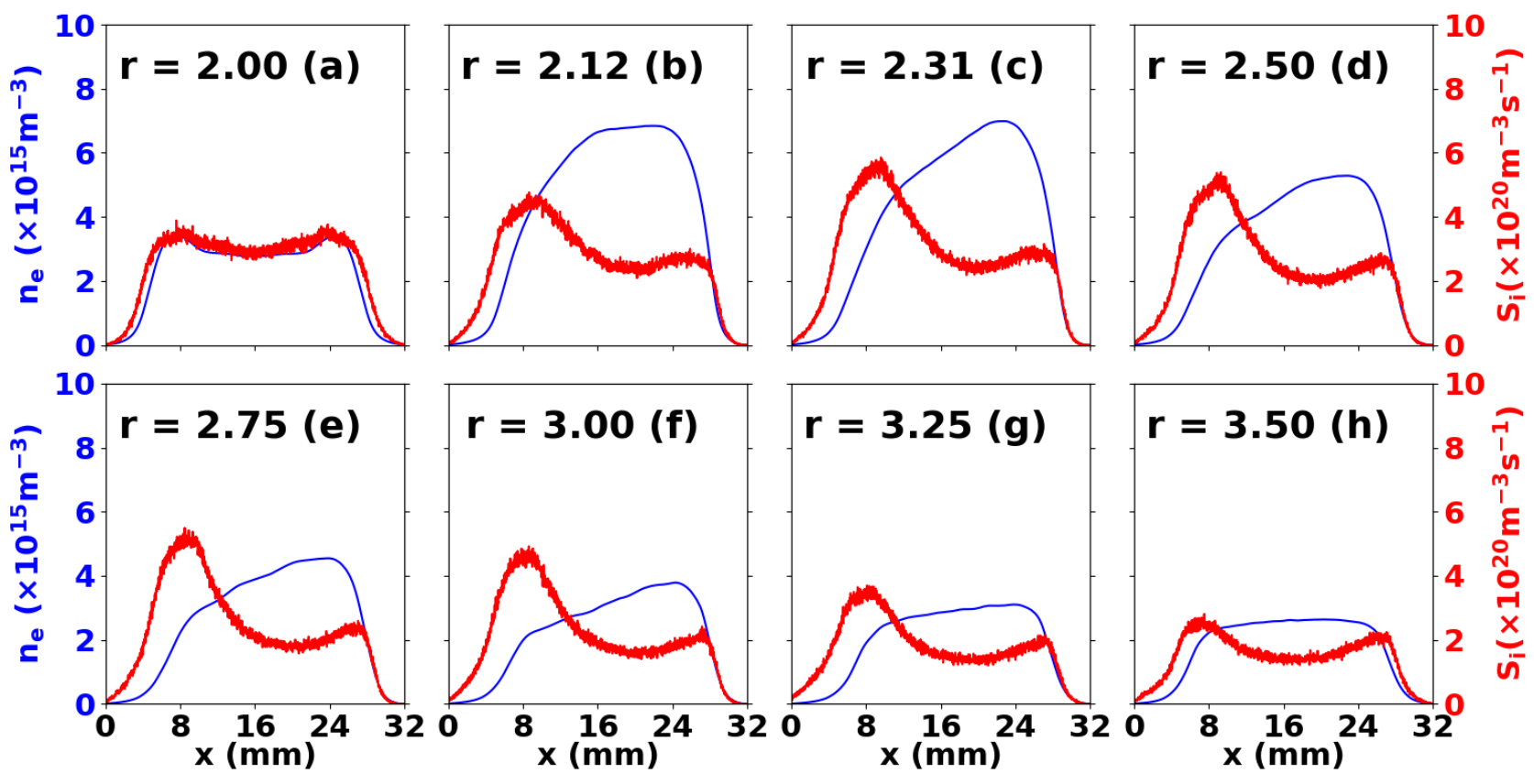} 
    \caption{
    Figure shows the spatial profiles of time-averaged electron density ($\mathrm{n_e}$, blue line, left y-axis) and time-averaged collisional ionization rate ($\mathrm{S_{i}}$, red line, right y-axis). As the magnetic field increases, both the electron density and ionization rate become increasingly asymmetric, with their maximum values occurring at $\mathrm{r = 2.31}$.}
    \label{ne_vol1}
\end{figure*}
\section{3.\ Results And Discussion}
Recall that in magnetized CCP discharges the first density enhancement appears for EBCR \cite{patil2020enhanced,patil2022electron} which corresponds to $\mathrm{r = 1}$ i.e. $\mathrm{f_{ce} = f_{rf}/2}$. Physically, this means that for $\mathrm{r = 1}$, the electrons complete a half-gyro-rotation during one RF cycle. 
As a result, electrons repeatedly gain energy by interacting with the oscillating sheath at specific phase of an RF cycle. Over time, this sustained energy gain enables them to reach ionization thresholds, leading to enhanced plasma density through electron-neutral collision-induced ionization.\\
In order to understand the physics of magnetized CCP beyond the EBCR, simulations are performed with an increasing applied magnetic field. The corresponding values $r$ in the present study are chosen such that $ \mathrm{2 \leq r \leq 3.5}$. The choice of the upper limit on $\mathrm{r}$ was dictated by the fact that beyond $\mathrm{r = 3.5}$, strong magnetization of the discharge causes a reduction in the loss of ions to the electrodes. The resulting density enhancement in this regime ($\mathrm{r \geq 3.5}$) through a MERIE-like mechanism has been extensively investigated in previous studies \cite{muller1989magnetically,lieberman1991model,hutchinson1995effects,park1997reactor,kushner2003modeling,vasenkov2004modeling}. Thus, with the above-mentioned range of $\mathrm{r}$, we aim to investigate the behavior of plasma when it transitions from mildly magnetized to strongly magnetized.
\subsection{3.1 \ Formation of asymmetric density profiles}
\begin{figure} 
    \centering
    \includegraphics[width=0.5\textwidth]{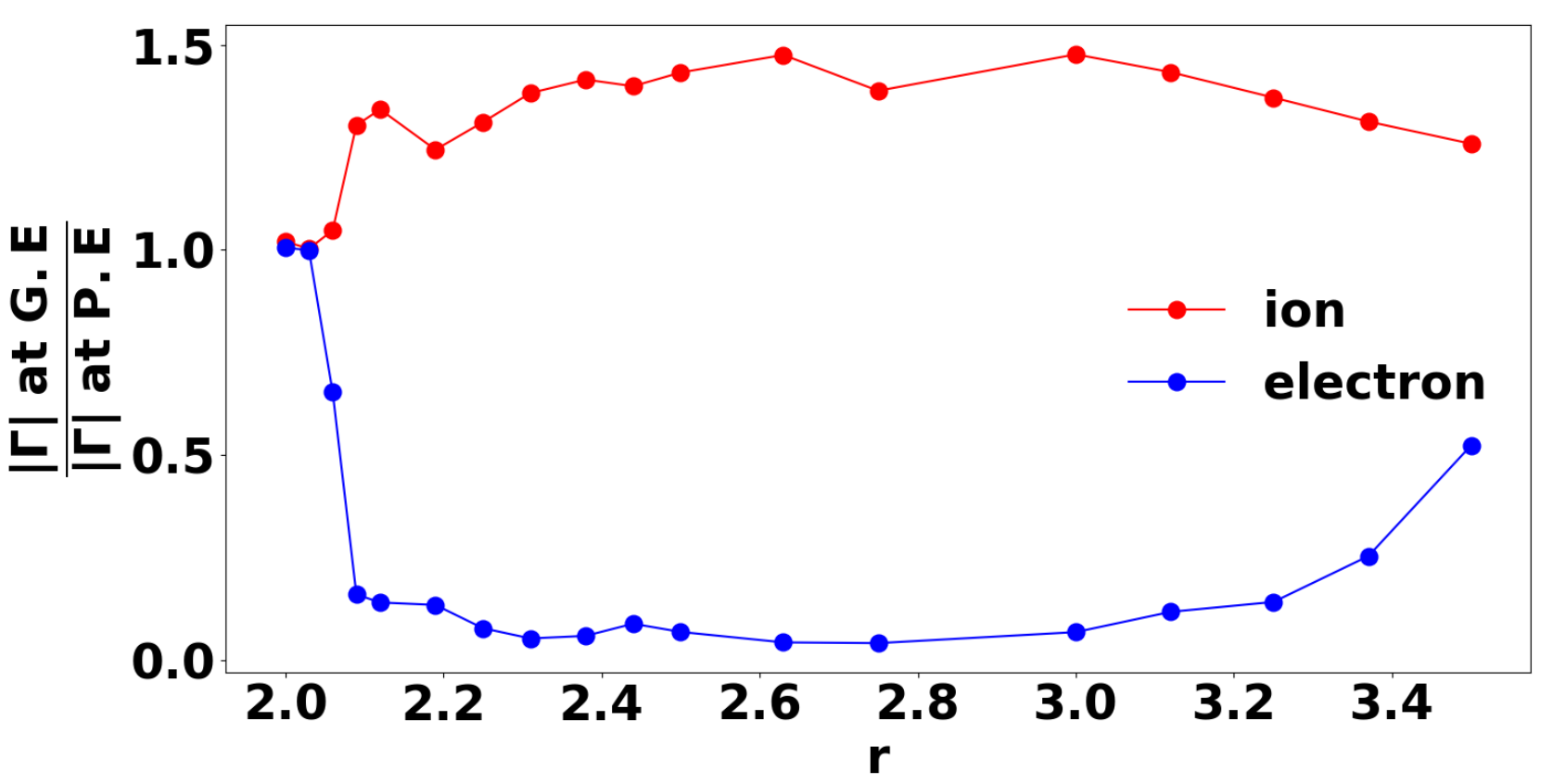}
    \caption{
    This figure shows the ratio of time-averaged ion and electron fluxes at the grounded electrode (G.E.) to those at the powered electrode (P.E.) as a function of magnetic field strength (denoted by $r$). The ion and electron flux ratios are represented by red and blue curves, respectively. In the magnetic field range $2.1 < r < 3.3$, corresponding to asymmetric discharges, the ion flux is higher at the G.E., while the electron flux is higher at the P.E.} 
    \label{flux_ratio}
\end{figure}

\begin{figure}  
    \centering
    \includegraphics[width=0.5\textwidth]{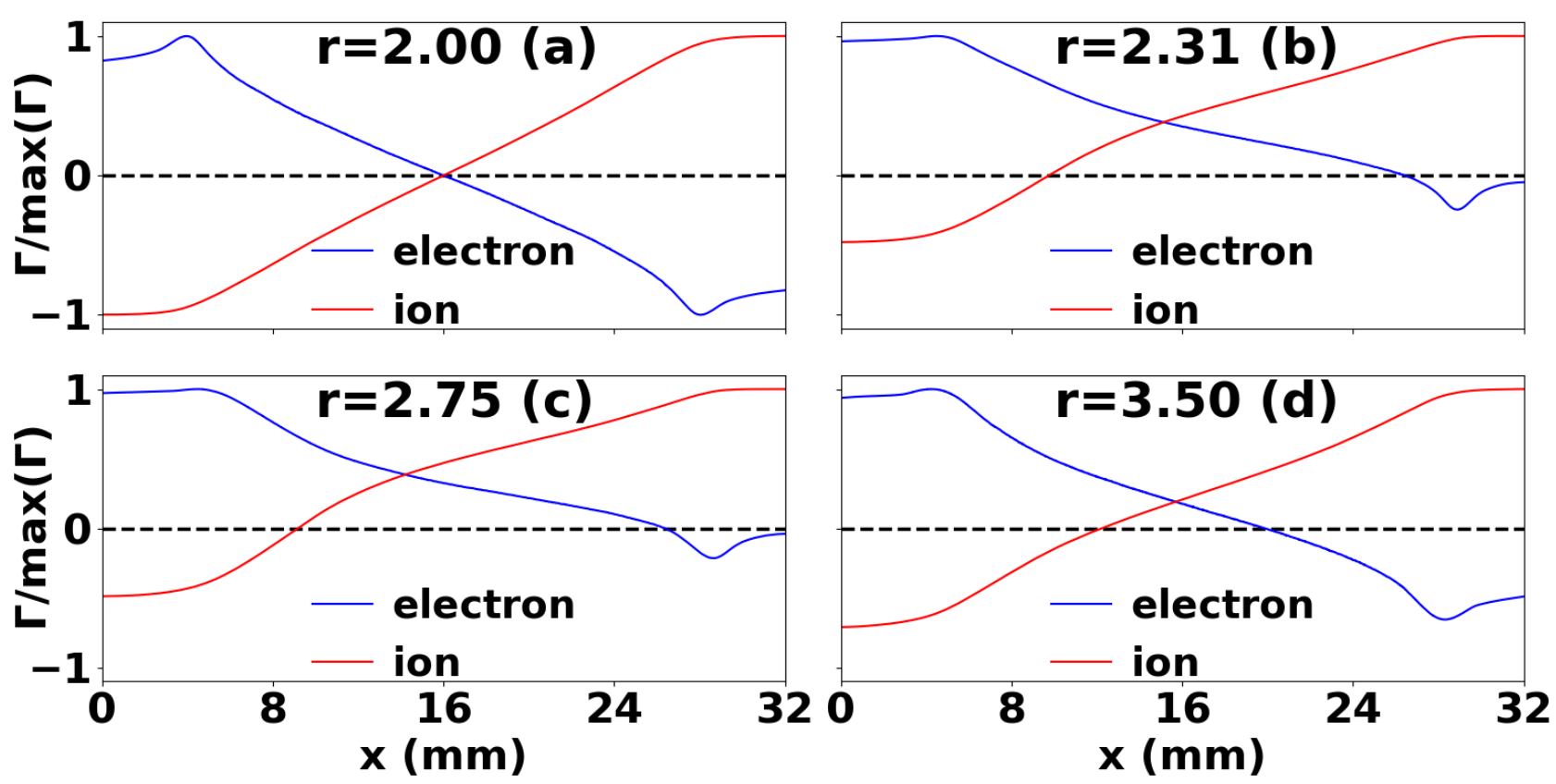}
    \caption{Time-averaged spatial profiles of ion (red) and electron (blue) fluxes for $r = 2.0, \, 2.31, \, 2.75, \, 3.5$). Both fluxes are normalized by their respective maximum values. In of asymmetric discharge ($r = 2.0$ and $3.5$), two fluxes are also nearly symmetric. For asymmetric discharges ($r = 2.31$ and $2.75$), ions are prominently lost from G.E. side and electrons are lost from P.E. side. Asymmetry in sheath near the two electrodes can be inferred from drop in the electron fluxes near the electrodes.}
    \label{current_density1}
\end{figure}

One of the prominent discharge characteristics found in the transition regime between $\mathrm{2.0 < r < 3.5}$, is the asymmetry in the spatial profile of the plasma density. As the magnetic field strength increases—or equivalently, as $\mathrm{r}$ increases—the plasma density profile transitions from symmetric to asymmetric and then returns to a symmetric form. This trend can be seen from Fig \ref{ne_vol1}, where the steady-state spatial profile of the time averaged electron number density (shown in blue) and ionization rate (shown in red) are plotted for various values of $\mathrm{r}$. Both quantities are averaged over the last 1000 RF cycles. For $\mathrm{r = 2.0}$ i.e. $\mathrm{f_{ce} = f_{rf}}$ (see Eq. (1)), both electron number density and ionization rate profiles are symmetric around the discharge center but peak near the sheath edge. The additional ionization near the edge of the sheath is mainly due to electron trapping in this region. As the magnetic field strength increases, the discharge becomes increasingly asymmetric. This growing asymmetry is accompanied by an increase in both the peak electron density and the ionization rate.
The figure shows that the highest degree of asymmetry occurs for $\mathrm{r = 2.31}$, where the peak electron number density is $\sim \mathrm{7 \times 10^{15}\,m^{-3}}$. As the magnetic field increases beyond r=2.31, the discharge begins to exhibit a gradual reduction in both asymmetry and the peak values of the electron density and the ionization rate. This trend continues until a symmetric discharge profile is restored at r=3.5 .\\
Interestingly, fig \ref{ne_vol1} also shows that the density profile peaks near the grounded electrode, while the ionization rate profile reaches its maximum near the powered electrode. This distinctive feature of asymmetric discharges has also been reported in recent work by Yao et al \cite{yao2024spontaneous}. This asymmetry was attributed to the differential influence of the magnetic field on electrons depending upon their energy. In addition, we observe that with growing asymmetry the magnitude of the ion flux is higher on the grounded electrode than on the powered electrode side. In order to maintain the quasi-neutrality of the bulk plasma, we expect a greater loss of electrons from the powered electrode side. This trend can be inferred from fig. \ref{flux_ratio}, where we have plotted the ratio of magnitude of fluxes at G.E. to P.E. for both ions (in red) and electrons(in blue) for changing magnetic field strengths ($\mathrm{r}$). Initially, when the discharge is symmetric ($\mathrm{r = 2}$), this ratio is close to unity for both ions and electrons due to the equal loss from both sides. However, with increasing $\mathrm{r}$, the ratio for ion fluxes exceeds unity, implying greater flux on the G.E. compared to P.E. The reverse trend is observed for the electron flux. Finally, when $\mathrm{r}$ approaches 3.5 both the ratios for ion and electrons get closer to unity. This shows that for asymmetric discharges, the ion loss is dominant from the G.E. side, and electrons are lost from the P.E. side. As the discharge gradually transitions toward a more symmetric state, the particle losses also become increasingly balanced between the two electrodes.\\
Figure \ref{current_density1} shows the spatial profiles of the time-averaged ion and electron fluxes for selected values of $r$, illustrating how the discharge evolves as r increases from 2.0 to 3.5.
This figure supports the trend represented in fig.\ref{flux_ratio}. Here, for both ion (red) and electron (blue) profiles, they are normalized by their respective maximum values. For $\mathrm{r = 2.0}$ (see Fig. 4(a)), when the discharge is symmetric, we observe that both ion and electron flux profiles are perfectly symmetric. In addition, the presence of an electrostatic sheath near the two electrodes can also be inferred from the drop in electron flux near $\mathrm{x = 0}$ and $\mathrm{32}$ mm. As the discharge becomes asymmetric ($\mathrm{r = 2.31}$ and $\mathrm{2.75}$), the ion and electron fluxes are highest at G.E. and P.E., respectively. The asymmetry in the sheath regions near the G.E. and P.E. is evident from the absence of a significant drop in electron flux near the P.E. side, as shown in Fig. 4(b) and (c). This indicates that electron losses at the G.E. are almost negligible when the discharge is strongly asymmetric. As $r$ increases to 3.5, the discharge becomes more symmetric, resulting in nearly identical electron and ion flux profiles at both electrodes (see Fig. 4 (d)). \\
Asymmetry in the sheaths near G.E and P.E. can be seen from the absence of a drop in the electron flux near the P.E. side (see Figs. 4(b) and (c)). It is clear that the electron loss at the G.E. is almost negligible when the asymmetry is dominant. For $\mathrm{r = 3.5}$, the discharge tends to become symmetric with nearly-identical electron and ion profiles.   \\ 
\begin{figure*}
    \centering
    \includegraphics[width=0.9 \textwidth]{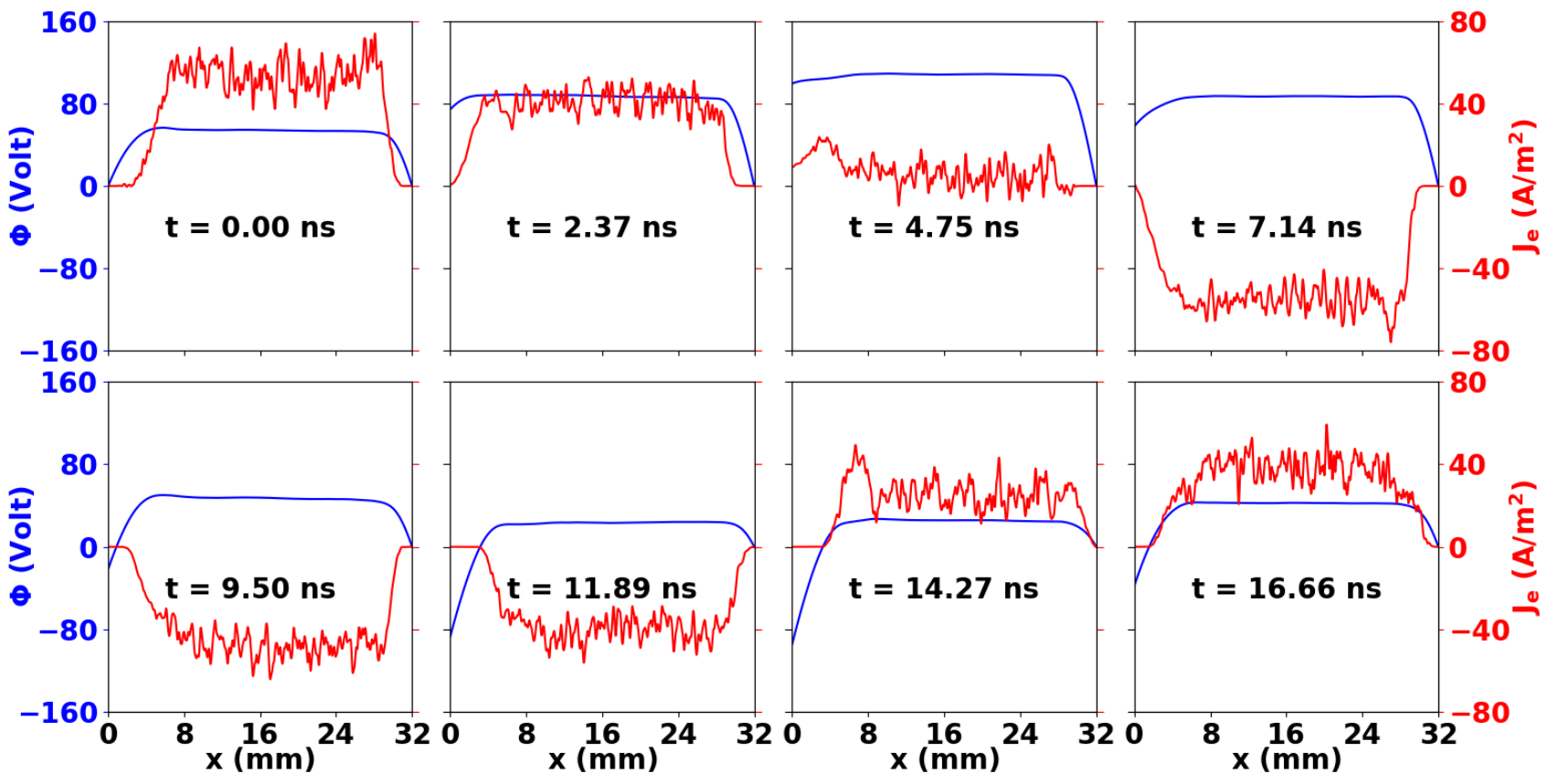}
    \caption{Snapshots of potential and electron current density for $\mathrm{r=2.31}$ for 1 rf cycle. For time interval between $2.37$ and $4.75$ ns, potential $\mathrm{\Phi}$ is almost flat at P.E. side. The excess loss of electrons during this period can be seen from positive spike in the electron current density $\mathrm{j_e}$. }
    \label{potential_snap}
\end{figure*}
The mechanism of loss of electrons prominently from the P.E. side, in case of asymmetric discharge, can be understood with the help of fig. \ref{potential_snap}. For this figure, we have considered the case with $\mathrm{r = 2.31}$ for which the discharge asymmetry is the most prominent. Here, instantaneous spatial profiles of the electrostatic potential ($\Phi$) and the electron current density ($\mathrm{j_e}$) are plotted for various times within one RF cycle, after the steady-state is achieved. The time $\mathrm{t = 0}$ ns corresponds to the moment when the applied RF voltage on the P.E. crosses zero. In the figures, the electron current density $\mathrm{j_e}$ is plotted, where positive values (shown in red) indicate electron flux directed toward the P.E., while negative values indicate flux toward the G.E.\\
For example, during the time interval from $\mathrm{t = 0}$ to $\mathrm{t = 4.75}$ ns, $\mathrm{j_e}$ remains positive, signifying that electrons are moving toward the P.E. During this phase, the sheath near the P.E. collapses, allowing electrons to flow into the electrode, while the sheath on the G.E. side becomes stronger, repelling electrons. In specific instances—namely $\mathrm{t = 2.37}$ ns and $\mathrm{t = 4.75}$  ns—we observe a near-complete collapse of the sheath at the P.E. This is evident from the electrostatic potential profile, which becomes nearly flat near $\mathrm{x = 0}$ mm, implying a vanishing electric field and a subsequent increase in electron loss from the P.E. side.\\
In contrast, during other time intervals such as $\mathrm{t = 7.14}$ ns, $\mathrm{t = 9.5}$ ns, and $\mathrm{t = 11.89}$ ns, the electron current density becomes negative, indicating that electrons are directed toward the G.E. However, the electrostatic potential near $\mathrm{x = 32}$ mm (G.E. side) never becomes entirely flat, suggesting that the sheath does not collapse completely, and hence, electron loss from this side remains minimal.\\
This time-dependent behavior leads to an important conclusion: in every RF cycle, the electrostatic potential dynamically adjusts itself to preferentially allow electron loss from the P.E. side. This asymmetry in electron loss plays a crucial role in maintaining the quasi-neutrality of the bulk plasma and results in a time-averaged potential drop across the discharge. As a consequence, ion and electron losses become asymmetric—ions are predominantly lost from one electrode (typically the G.E.), while electrons are lost from the other (P.E.). This asymmetry in particle losses ultimately determines the spatial profile of the steady-state plasma density, which peaks on the side with dominant ion loss.

\begin{figure*}
    \centering
    \includegraphics[width=0.9 \textwidth]{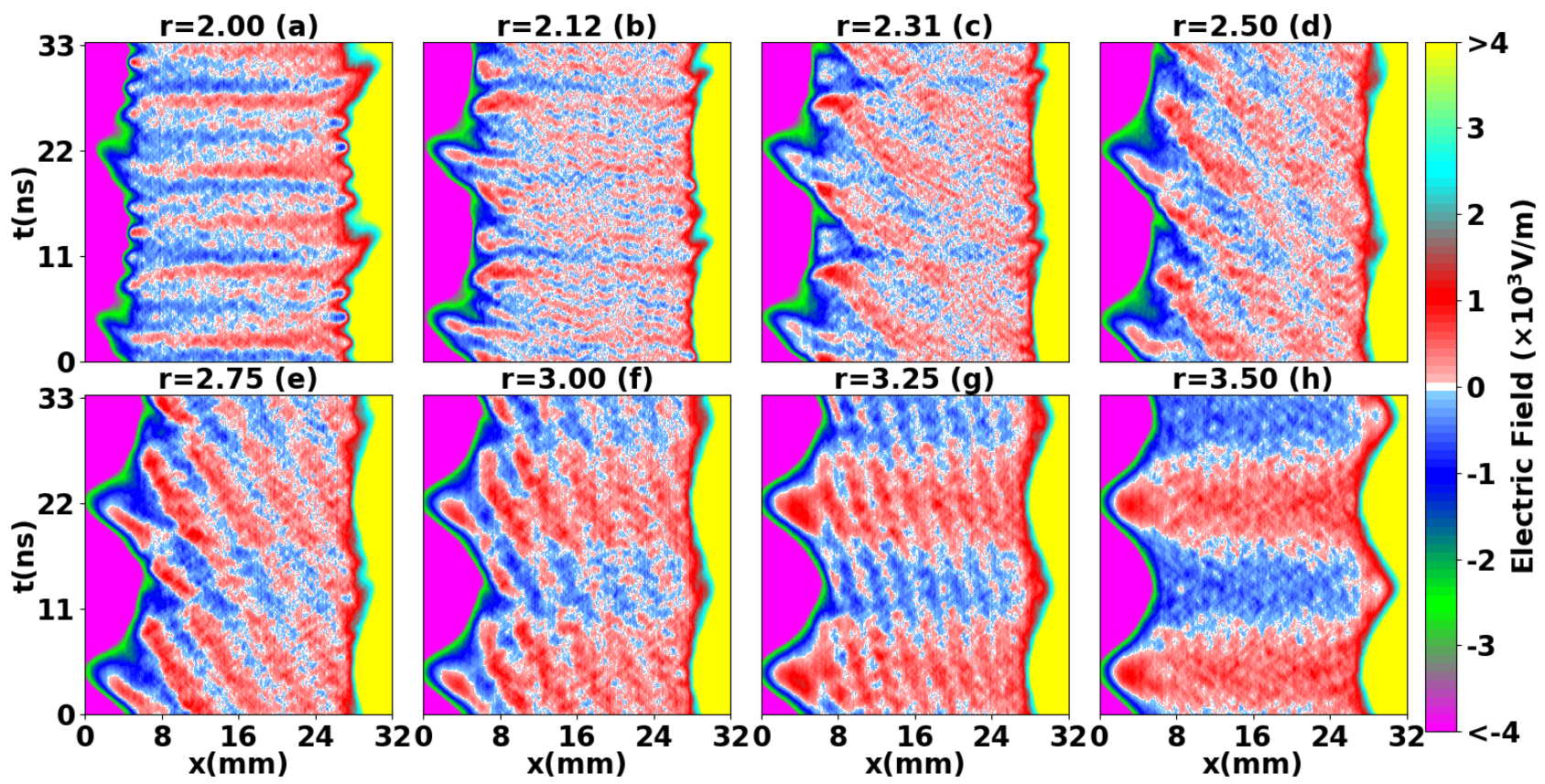} 
    \caption{Spatiotemporal evolution of the electric field over last two RF cycles after reaching steady state at different magnetic field strengths.}  
    \label{electric_field}
\end{figure*}
\subsection{3.2 \  Bernstein Wave Dynamics in the Bulk Plasma}
\begin{figure}
    \centering
    \includegraphics[width=0.5 \textwidth]{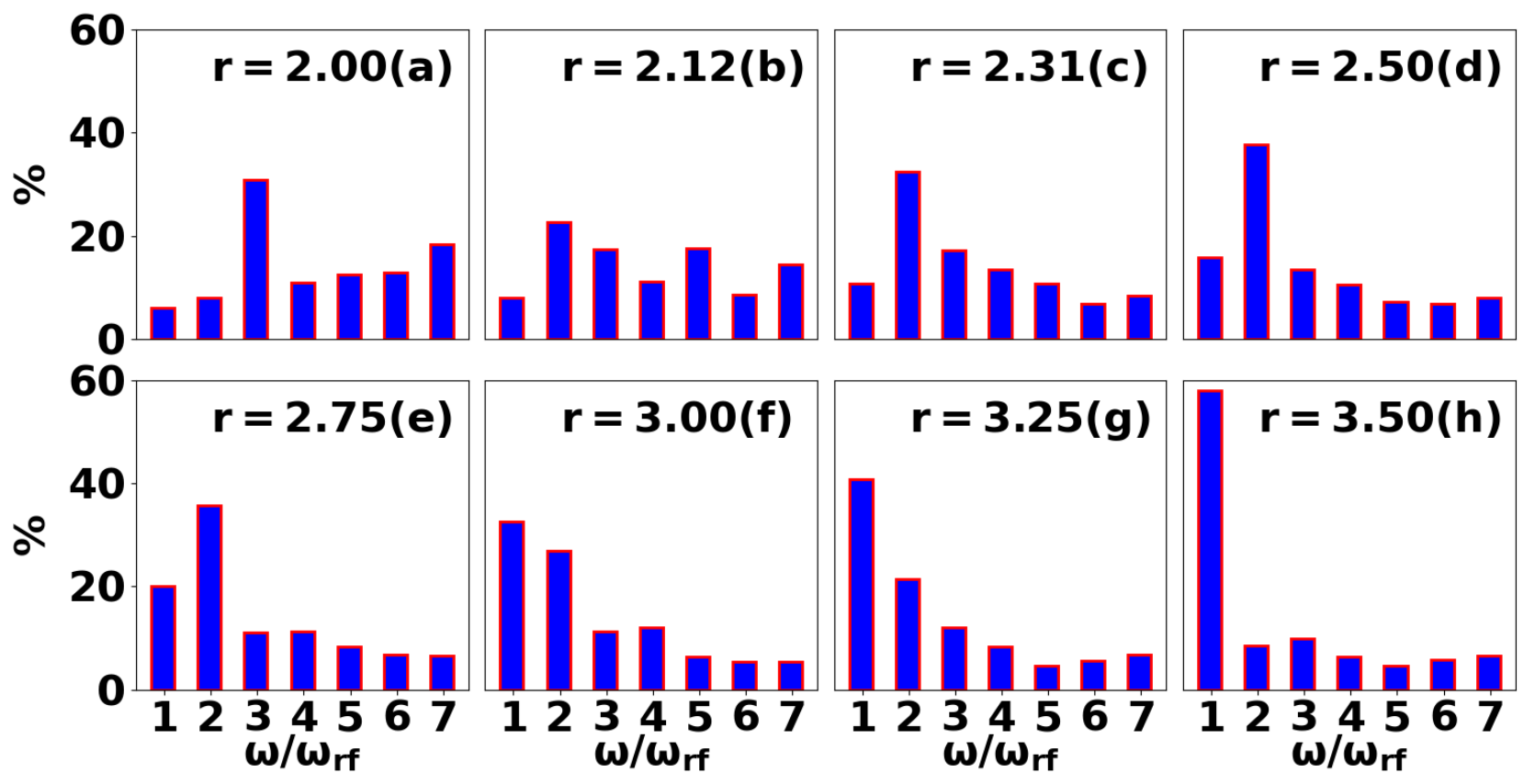}
    \caption{Temporal FFT analysis of electric field inside bulk region from $\mathrm{8 \ mm}$ to $\mathrm{24}$ mm.}
    \label{fft_bar}
\end{figure}

\begin{figure*}
\centering
\includegraphics[width=0.9 \textwidth]{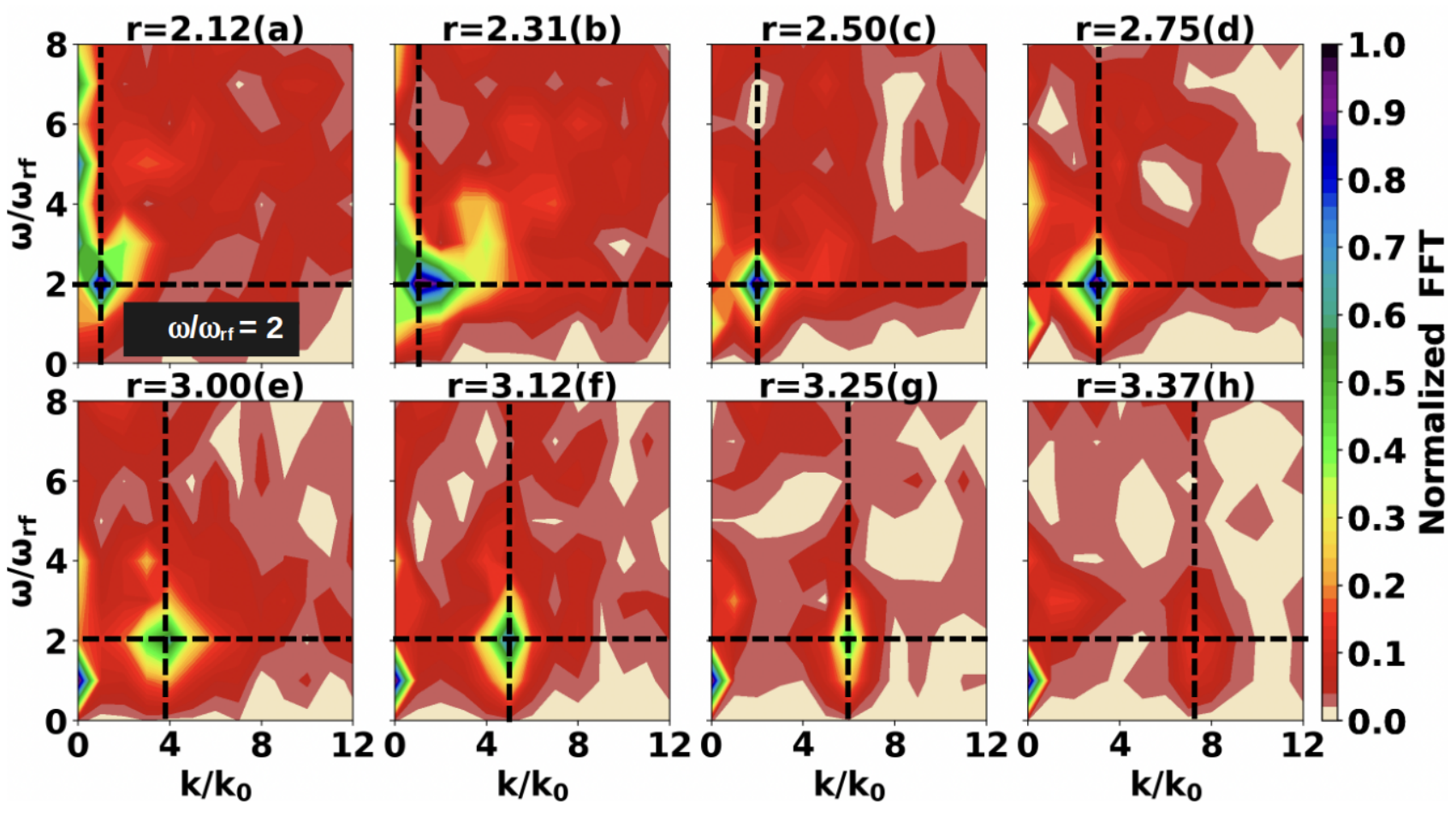} 
\caption{Normalized 2D FFT magnitude spectrum of the electric field data, illustrating the distribution of frequencies and their corresponding normalized magnitudes. The dotted  horizontal and vertical lines indicate the second harmonic frequencies($\mathrm{\omega/\omega_{rf}=2}$) and  associated normalized wave number($\mathrm{k/k_0}$).}
\label{2d_fft}
\end{figure*}

One of the fundamental challenges in understanding CCP discharge lies in the nature of electric field penetration into the bulk plasma. Under typical conditions, the plasma frequency in the bulk is substantially higher than the applied RF. As a result, the applied electric field remains largely confined within the sheath regions adjacent to the electrodes, whereas only a small portion penetrates the quasi-neutral core of the plasma.\\
Despite this confinement, the transient electric field that does manage to penetrate the bulk plays a crucial role in plasma heating and sustenance. Therefore, a thorough understanding of the physical mechanisms governing electric field penetration into the bulk plasma is essential for advancing the physics of CCP discharges, especially under low-pressure or magnetized conditions.\\
Previous studies, notably those of Kaganovich et al. \cite{kaganovich2006revisiting}, have proposed that penetration of the RF field into the bulk can occur through kinetic mechanisms. In the current study, we build upon this concept and demonstrate that, within the transitional magnetic field regime of interest, the RF field penetrates the plasma in the form of EBWs. These waves, which are electrostatic in nature and capable of propagating in overdense plasmas, provide a viable channel for energy transport deep into the plasma bulk, where conventional electromagnetic waves would otherwise be evanescent.
This insight advances the theoretical understanding of RF energy coupling in magnetized CCPs.\\
Bernstein waves are electrostatic waves traveling perpendicular to the external magnetic field. They are resonant modes excited at the harmonics of the cyclotron frequency of the charged species inside the plasma. Because these modes are electrostatic in nature, they can propagate inside the overdense plasma without any restriction on cutoff frequency. In fact, this property makes these waves a potential candidate for heating of the bulk plasma. The excitation of these waves has been shown to be facilitated by the presence of density gradients inside the plasma \cite{sugaya1971parametric,ram2000excitation}. Therefore, the present operation regime, having gradients in plasma density and  $\mathrm{f_{rf} \sim f_{ce}}$, is ideal for the excitation of electron Bernstein waves. \\    
To demonstrate the formation of these waves, we look at the spatiotemporal profile of the electric field over two RF cycles as shown in  Figure~\ref{electric_field}. Since we are interested in the behavior of electric field penetrating into the bulk plasma, the color scale of this plot is saturated for the magnitude of electric fields beyond $4\times{10}^3$ V/m, a typical value at the edge of the sheath. The maximum magnitude of the electric field of the sheath is found to be around $2\times{10}^4$ V/m. Therefore, the chosen color scale allows us to clearly demonstrate the distinct features of electric fields within the bulk plasma.
These RF-cycles are chosen after the steady state is reached. With increasing magnetic field strength, transient electric field structures begin to penetrate more prominently into the bulk plasma region. At $\mathrm{r = 2.0}$, shown in Fig.~\ref{electric_field}(a), the electric field structures within the bulk plasma appear parallel to the $\mathrm{x}$-axis. As $\mathrm{r}$ increases from 2.12 (Fig.~\ref{electric_field}(b)) to 2.75 (Fig.~\ref{electric_field}(e)), these structures become prominent and more inclined. The inclination of these field structures in the $x-t$ plane is indicative of the existence of an electrostatic wave traveling inside the bulk plasma. In particular, the speed of these waves has an inverse dependence on the slope of these structures in the spatial-temporal plots. It can be seen that beyond $\mathrm{r = 2.31}$, the speed of these waves inside the bulk plasma decreases with increasing magnetic field strengths. It should also be noted that the peak density in this transition regime was maximized at $\mathrm{r = 2.31}$. 
With increasing magnetic field strength, a noticeable reduction in the wavelength of the inclined electric field structures is observed in the spatiotemporal plots. This behavior is consistent with the theoretical expectation that the phase velocity of electrostatic waves, such as EBWs, decreases as the magnetic field increases. Since phase velocity is directly related to wavelength ($\mathrm{v_p = \omega/k}$), a decrease in phase velocity at a fixed frequency implies a shorter wavelength. This trend confirms that the wave structures observed within the plasma bulk are indeed influenced by the magnetic field, and their properties evolve accordingly. In addition, beyond $\mathrm{r = 2.75}$, as seen in Fig.~\ref{electric_field}(f) to (h), the inclined structures begin to break and eventually at $\mathrm{r = 3.5}$, they merge into thicker, nearly parallel large-scale structures. \\ 
\hspace*{0.3 cm} To characterize these waves, we performed a fast temporal Fourier analysis (FFT) of the electric field in the bulk plasma region ($\mathrm{8\,mm - 24\,mm}$). The results are presented in Figure~\ref{fft_bar}. At each position within this region, the FFT is computed, and the harmonic contributions are integrated. The figure displays this spatially integrated FFT of electric field, sampled over one RF-cycle. The horizontal axis denotes the frequency normalized to the applied RF frequency ($\mathrm{f_{rf} = 60\,MHz}$), while the vertical axis represents the percentage contribution of various harmonics in the bulk region. From the figure, it is evident that at $\mathrm{r = 2.0}$, the third harmonic is dominant, contributing approximately 30\%. As $\mathrm{r}$ increases, the contribution of the third harmonic decreases, at $\mathrm{r = 2.12}$, it reduces to around 19\%, while the second harmonic rises to nearly 25\%. The second harmonic reaches its maximum contribution of about 40\% at $\mathrm{r = 2.5}$. Beyond this point, the fundamental (first) harmonic becomes dominant, attaining a maximum contribution of nearly 60\% at $\mathrm{r = 3.5}$, corresponding to the complete merging of electric field structures (see Fig.~\ref{electric_field}(h)). Thus, as the magnetic field strength increases, we observe a systematic transformation in the harmonic content of the electric field fluctuations. Specifically, the higher-order harmonics gradually diminish, giving way to a dominant presence of lower harmonics. In particular, the second harmonic emerges as the dominant mode within the magnetic field ratio range $\mathrm{2.12 < r < 2.75}$, precisely where the plasma density profile becomes asymmetric. These observations suggest a strong correlation between magnetic field strength, harmonic content, and plasma asymmetry, shedding light on the mechanisms driving wave excitation and energy transport in magnetized CCP discharges. This is expected as the loss of wave-momentum depends on the density gradient, as argued in a companion paper submitted to Physical Review Letters \cite{gautam2025resonantly}.
Also, recall that with increasing magnetic field, in Fig.~\ref{electric_field} we observed shorter wavelength structures in the bulk plasma. Thus, cascading of higher harmonics into lower ones with increasing wavenumber is suggestive of the presence of the Bernstein wave inside the bulk plasma.\\
 In order to confirm the presence of Bernstein waves inside the bulk plasma, it is essential to construct its dispersion relation. We obtain the dispersion curve by taking the two-dimensional FFT of the spatial-temporal electric field, shown in Fig.~\ref{electric_field}. Here, we consider those $\text{r}$ values for which we observe a distinct tilt in the spatio-temporal electric field, indicating the presence of EBW. For 2D FFT analysis, we consider only the field present inside the bulk of the plasma, which is defined as the region between $\mathrm{8 \ mm \ to \ 24 \ mm}$. In Fig. ~\ref{2d_fft}, the magnitude of spectrum obtained by taking 2D FFT of the electric field inside the bulk plasma is shown. Here, the frequencies are normalized with the applied RF frequency, i.e., $\mathrm{f_{rf}=60 \ MHz}$, and the wave numbers $\mathrm{k}$ are normalized with the inverse of the bulk plasma length, $\mathrm{k_{0}=1/0.016 \approx 62.5 \ m^{-1}}$. Specifically, we are interested in the wavenumber ($k$) associated with the second harmonic of the RF frequency, since this is the fundamental mode of the electron Bernstein wave in the long-wavelength limit. In this figure, the horizontal dotted line marks the second harmonic ($\mathrm{f/f_{rf} = 2}$), while the vertical dotted line indicates the associated normalized wave number ($\mathrm{k/k_{0}}$). The intersection point of these two lines gives us the frequency and wave-number associated with the wave for a given magnetic field ($r$ value). Thus, using each of these intersection points for various $r$ values, we can construct the dispersion relation of the plasma wave seen in Fig.~\ref{electric_field}. \\   
Theoretically, the relation between frequency ($\omega$) and wavenumber ($\mathrm{k_{\perp}}$) for an EBW \cite{bellan2008fundamentals} is obtained by  
\begin{equation}
\label{e_disp}
\mathrm{1= \dfrac{2e^{-\lambda}}{\lambda} \sum_{n=1}^{\infty} \dfrac{n^{2}\omega_{pe}^{2}}{\omega^{2}-n^{2}\omega_{ce}^{2}} \text{I}_{\text{n}}(\lambda)}.
\end{equation}
This dispersion relation applies to the electrostatic case ($k_{\parallel} = 0$), assuming cold ions and a Maxwellian electron population with temperature $T_e$. The spatial profiles of both the plasma density and the magnetic field are assumed to be uniform. The relevant timescales are characterized by the inverse plasma frequency ($\omega_{\text{pe}}^{-1}$) and the inverse electron cyclotron frequency ($\omega_{\text{ce}}^{-1}$). \\
The assumption of uniform plasma density remains valid as long as the density scale length is significantly larger than the wavelength of the EBWs. This condition is satisfied in our case, since long-wavelength modes—comparable to the size of the bulk plasma—cannot propagate due to the finite system size. Therefore, $\omega_{\text{pe}}^{-1}$ in the dispersion relation is evaluated using the average plasma density in the bulk region.

\begin{figure}
    \centering
    \includegraphics[width=0.5 \textwidth]{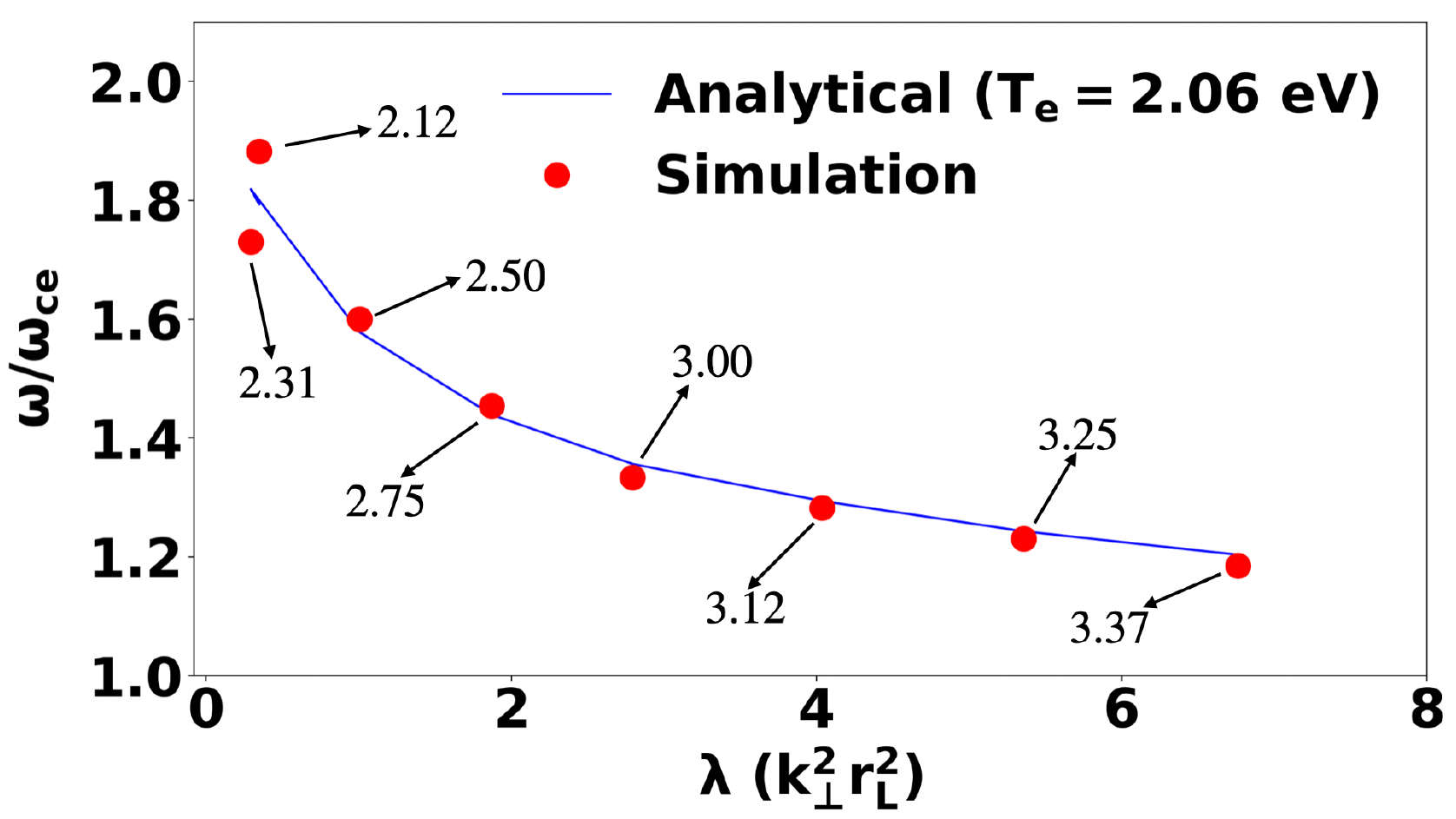} 
    \caption{Electron Bernstein Wave Dispersion Relation. The figure illustrates the dispersion characteristics of the dominant frequency, which transitions from the second harmonic ($\mathrm{2 \  \omega_{ce}}$) toward the fundamental harmonic of cyclotron frequency ($\mathrm{\omega_{ce}}$) as $\mathrm{k_{\perp}^{2}r_{L}^{2}}$ increases. The solid red circles represent the results obtained from numerical simulations, while the blue curve corresponds to the analytical dispersion relation.} 
    \label{bernstein_dr}
\end{figure}

\begin{figure}
    \centering
    \includegraphics[width=0.5 \textwidth]{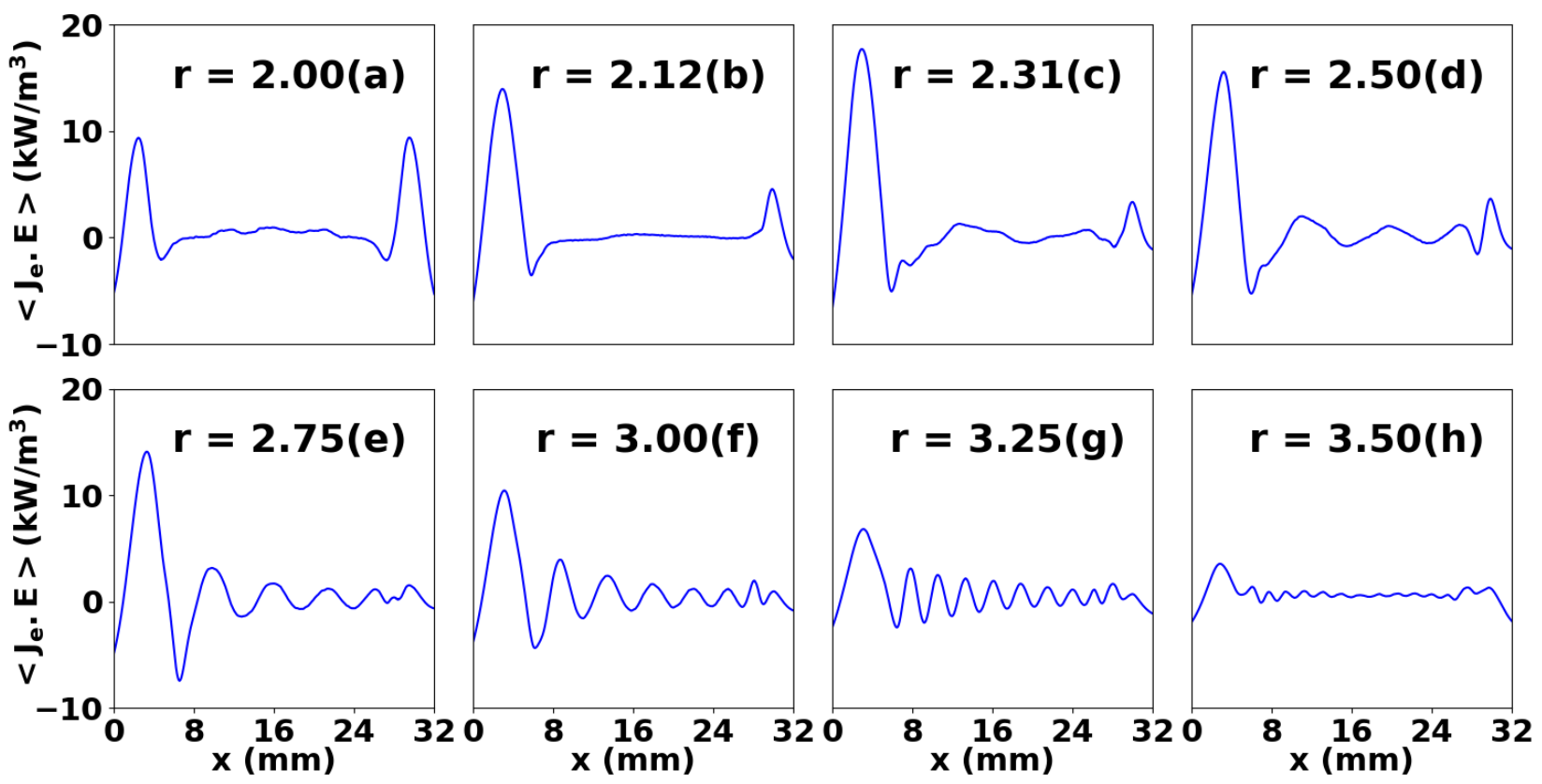} 
    \caption{Time average electron heating ($\langle \mathbf{j}_e \cdot \mathbf{E}\rangle$) with increasing magnetic field. The presence of wiggles in the bulk plasma demonstrates the contribution of the Bernstein waves to its heating. In case of asymmetric discharges ($2.12 \leq r \leq 3.25$), the heating at the powered electrode dominates over the grounded electrode. }
    \label{e_heating}
\end{figure}

\begin{figure}
    \centering
    \includegraphics[width=0.5 \textwidth]{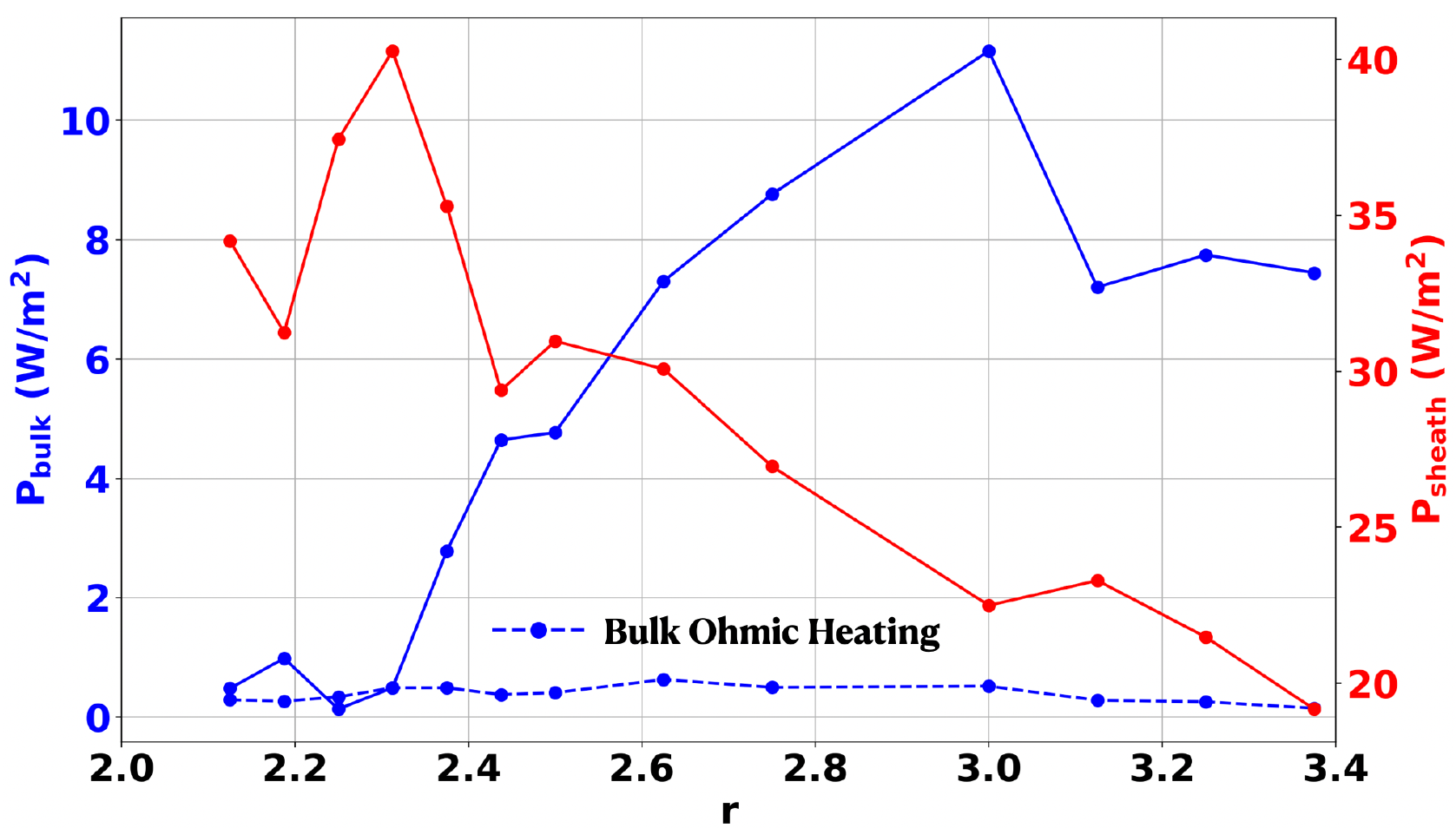} 
     \caption{The total electron heating in the bulk ($P_{\text{bulk}}$), sheath regions ($P_{\text{sheath}}$), and the ohmic heating in the bulk, as defined in Eq.~\ref{eq:bulk_sheath_heating}, are plotted as a function of $r$. For $r > 2.1$, where the presence of electron Bernstein waves (EBWs) is observed, a significant increase in bulk heating (blue curve) is evident. The ohmic heating within the bulk (blue dotted curve) remains negligible in this region. Correspondingly, the sheath heating (red curve) decreases. This indicates that the enhanced bulk heating is primarily due to EBW-induced wave heating.} 
     
    \label{5mtorr_heat}
\end{figure}

\begin{figure*}
    \centering
    \includegraphics[width= 0.9 \textwidth]{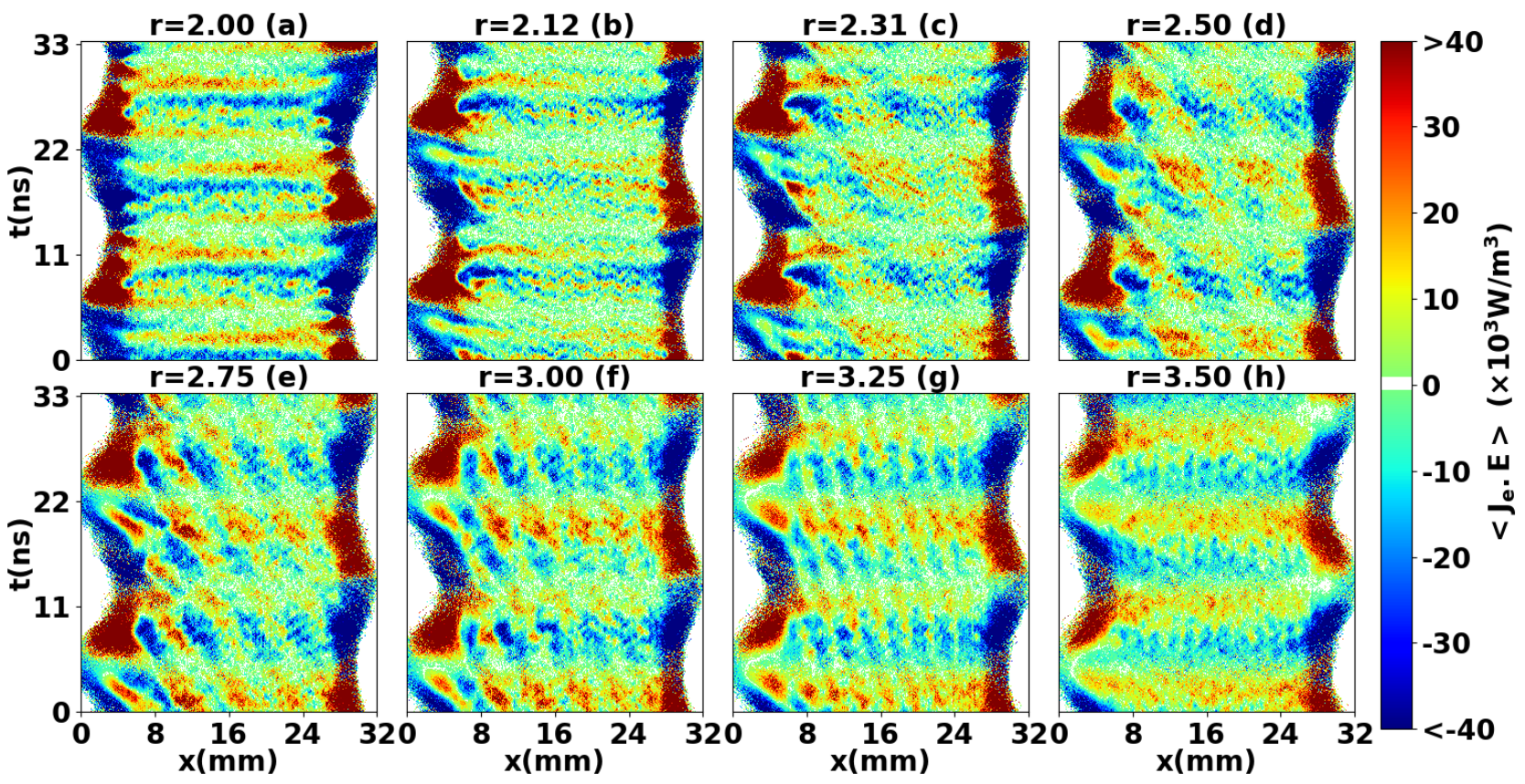} 
    \caption{Figure demonstrates the Spatial temporal electron heating at different magnetic field. The structures seen here show the wave propagation in bulk , which is consistent with spatial temporal electric field data shown in fig \ref{electric_field}.} 
    \label{e_heating_2d}
\end{figure*}

Also, $\text{I}_{\text{n}}$ is the modified Bessel function of order $\text{n}$ and $\lambda$ is the Larmor parameter defined as
\begin{equation}
\label{lambda}
\mathrm{\lambda =k_{\perp}^{2}  r_{L}^{2}}.
\end{equation}
Note that $\lambda$ scales with the
square of the inverse of the wavelength normalized by the electron Larmor
radius $\mathrm{r_{L}}$.
The Larmor radius is calculated by using the expression as follows:
\begin{equation}
    \label{larmor}
    r_L=0.0238 \sqrt{T_e}/B 
\end{equation} 
Here, $r_L$ is expressed in meters, $T_e$ in electron volts (eV), and 
$B$ in Gauss. The electron temperature $T_e$ is determined from the statistical average of the kinetic energy associated with the fluctuating (thermal) component of the particle velocities of the macro-particles in the bulk plasma.\\
For the fundamental harmonic, we consider only the first two terms in the series expansion of Eq.~\ref{e_disp}, which gives the following.
\begin{equation*}
\begin{split}
1 &= \dfrac{2e^{-\lambda}}{\lambda} \left[ \dfrac{\omega_{pe}^{2}}{\omega^{2}-\omega_{ce}^{2}} \text{I}_{1}(\lambda) + \dfrac{4 \omega_{pe}^{2}}{\omega^{2}-4 \omega_{ce}^{2}} \text{I}_{2}(\lambda) \right].
\end{split}
\end{equation*}
Solving quadratic in $\omega^2$, we get 

\begin{equation}
\label{neg_sol}
\frac{\omega^2}{\omega_{ce}^2} = \dfrac{5+\alpha}{2} - \dfrac{1}{2} \sqrt{9 + 10\alpha + \alpha^{2} - 16\beta}.
\end{equation}
Here,
\begin{equation*}
\begin{split}
\alpha &= \dfrac{2e^{-\lambda}}{\lambda} \dfrac{\omega_{pe}^{2}}{\omega_{ce}^{2}} \left[ \text{I}_{1}(\lambda) + 4 \ \text{I}_{2}(\lambda) \right], \\
\beta &= \dfrac{2e^{-\lambda}}{\lambda} \dfrac{\omega_{pe}^{2}}{\omega_{ce}^{2}} \left[ \text{I}_{1}(\lambda) + \text{I}_{2}(\lambda) \right]
\end{split}
\end{equation*}
Note that asymptotically, in the long-wavelength limit ($\lambda \rightarrow 0$), we get $\mathrm{\alpha \simeq \beta \simeq \dfrac{\omega_{pe}^{2}}{\omega_{ce}^{2}}>>1}$ and $\mathrm{\omega \simeq 2 \omega_{ce}}$. On the other hand, in the short wavelength limit ($\mathrm{\lambda \rightarrow \infty}$), Eq.~\ref{neg_sol} reduces to $\mathrm{\omega \simeq \omega_{ce}}$. The dependence of $\mathrm{\omega/\omega_{ce}}$ on $\mathrm{\lambda = \left(k_{\perp}r_L\right)^2}$ for the fundamental harmonic ($\mathrm{n = 1}$) of EBW, obtained from Eq.~\ref{neg_sol} is plotted in Fig.~\ref{bernstein_dr}. In this figure, the blue line represents the analytical solution, whereas the solid red circles are the numerically obtained modes from the 2D FFT analysis, presented in Fig.~\ref{2d_fft}. Recall that for the EBW mode obtained in simulations, it corresponds to $\mathrm{f/f_{rf} = 2}$ and its associated wavenumber is obtained from the intersection of black-dotted lines as shown in Fig.~\ref{2d_fft}. We can see that the fundamental EBW mode, obtained in the simulations (solid red circle), is in excellent agreement with the analytical curve (blue line). This confirms that the wave-like structure observed in the presence of asymmetric discharge  is indeed a fundamental EBW mode.\\       
In the context of fusion plasmas, the use of Bernstein waves for plasma heating is extensively studied \cite{ram2000excitation}. The existence of these waves, as demonstrated above, also opens similar opportunities in CCP discharges. We observe distinctive features of heating of the bulk plasma by these waves in the time-averaged electron heating ($\langle \mathbf{j}_e \cdot \mathbf{E}\rangle$) presented in Fig.~\ref{e_heating}. In this figure, the wavelength of wiggles in $\langle \mathbf{j}_e \cdot \mathbf{E}\rangle$ inside the bulk plasma decreases with increasing magnetic field. This is consistent with the increasing wavenumber $k$ associated with the Bernstein wave, reported in Fig.~\ref{electric_field}. This figure also demonstrates the correlation between the discharge asymmetry and electron heating. For example, at $r = 2.0$, where the discharge remains symmetric, the heating of electrons near both electrodes is identical, indicating balanced power absorption on either side. However, in the case of asymmetric discharges ($r > 2.0$), as illustrated in Fig.~\ref{e_heating}(b)-(f), electron heating becomes significantly more pronounced near the powered electrode compared to the grounded electrode. The enhanced electron heating near the powered electrode in the case of asymmetric discharges is due to the increased electron flux on the powered electrode side compared to the grounded electrode side. From the spatial profile of the time-averaged electron heating, we calculate total heating (per unit area), inside the bulk plasma ($P_{\text{bulk}}$) and in the sheath region ($P_{\text{sheath}}$). Additionally, we estimate steady-state ohmic heating inside the bulk ($P_{bulk \ ohmic}$) to demonstrate the role of wave heating inside this region. These quantities are defined as follows:
\begin{equation}
    \begin{split}
        P_{\text{sheath}} &= \int_0^{x_{sp}} \langle j_e \cdot E\rangle dx' + \int_{x_{sg}}^L \langle j_e \cdot E\rangle dx', \\\\
         P_{\text{bulk}} &= \int_{x_{sp}}^{x_{sg}} \langle j_e \cdot E\rangle dx' , \\\\
          P_{bulk \ ohmic}&= \int_{x_{sp}}^{x_{sg}} \sigma \langle E^{2}\rangle dx' 
       \end{split}
    \label{eq:bulk_sheath_heating}
\end{equation}
Here, $L$ is the discharge length, while $x_{sp}$ and $x_{sg}$ denote the sheath-edges at the powered and grounded electrodes, respectively. In our case, the bulk plasma size ($x_{sg} - x_{sp}$) is approximately 16 mm. Additionally, the contribution of ohmic heating inside the bulk plasma is estimated using steady-state conductivity ($\sigma$) expressed as $\sigma = \dfrac{n_{e} e^2}{m_e} \left(\dfrac{\nu_{en}}{\nu^{2}_{en}+\omega^{2}_{ce}}\right)$. Here, a simulated steady-state profile is used for the electron density $n_e$ and the maximum electron-neutral cross-section is used for the computation of the electron-neutral collision frequency $\nu_{en}$. The variation of the quantities defined in Eq.~\ref{eq:bulk_sheath_heating} as a function of $r$ is shown in Fig.~\ref{5mtorr_heat}. The figure shows that the total bulk heating (solid blue line) increases for $r > 2.1$, where the presence of EBWs is observed in the form of `wiggly' structures in the bulk plasma (see Fig.~\ref{e_heating}). This is due to the fact that the positive half-cycles of the oscillatory structures (“wiggles”) dominate over the negative half-cycles. In contrast, the ohmic heating within the bulk plasma (blue dotted line) remains significantly smaller compared to the total bulk heating, indicating collisionless heating of bulk electrons. In the same range of $r$ values ($r > 2.1$), the sheath heating (red curve) decreases. Typically, we observe that no more than 30$\%$ of the total electron heating ($P_{\text{sheath}} + P_{\text{bulk}}$) is deposited in the bulk plasma. Considering that sheath heating is predominantly due to stochastic heating mechanism, we expect that the overall contribution of stochastic heating is diminishing with increasing $r$ values. This suggests that the increase in the bulk heating observed in the figure is pre-dominantly due to wave heating by EBWs.
In order to conclusively demonstrate that the oscillatory features, or `wiggles', in the time averaged electron heating, are due to the presence of Bernstein waves, we look at the spatio-temporal variation of $\mathbf{j}_e \cdot \mathbf{E}$ over two rf-cycles, as shown in Fig.~\ref{e_heating_2d}. This figure reveals a propagation pattern closely mirroring that of the electric field structure shown in Fig.~\ref{electric_field}. As Bernstein waves traverse the bulk plasma, they modulate the local electric field, leading to synchronous oscillations in $\mathbf{j}_e \cdot \mathbf{E}$. In particular, these wave-like structures exhibit a decreasing wavelength with increasing magnetic field, consistent with the increasing wavenumber $k$ of the Bernstein waves. The spatial pattern of oscillations in the time-averaged electron heating aligns well with this trend, providing strong evidence that these wiggles are indeed a manifestation of Bernstein wave activity within the bulk plasma.

\subsection{3.3 \quad Analysis of Electron Trajectories in case of Asymmetric discharges}
\begin{figure*}
    \centering
    \includegraphics[width=1.0 \textwidth]{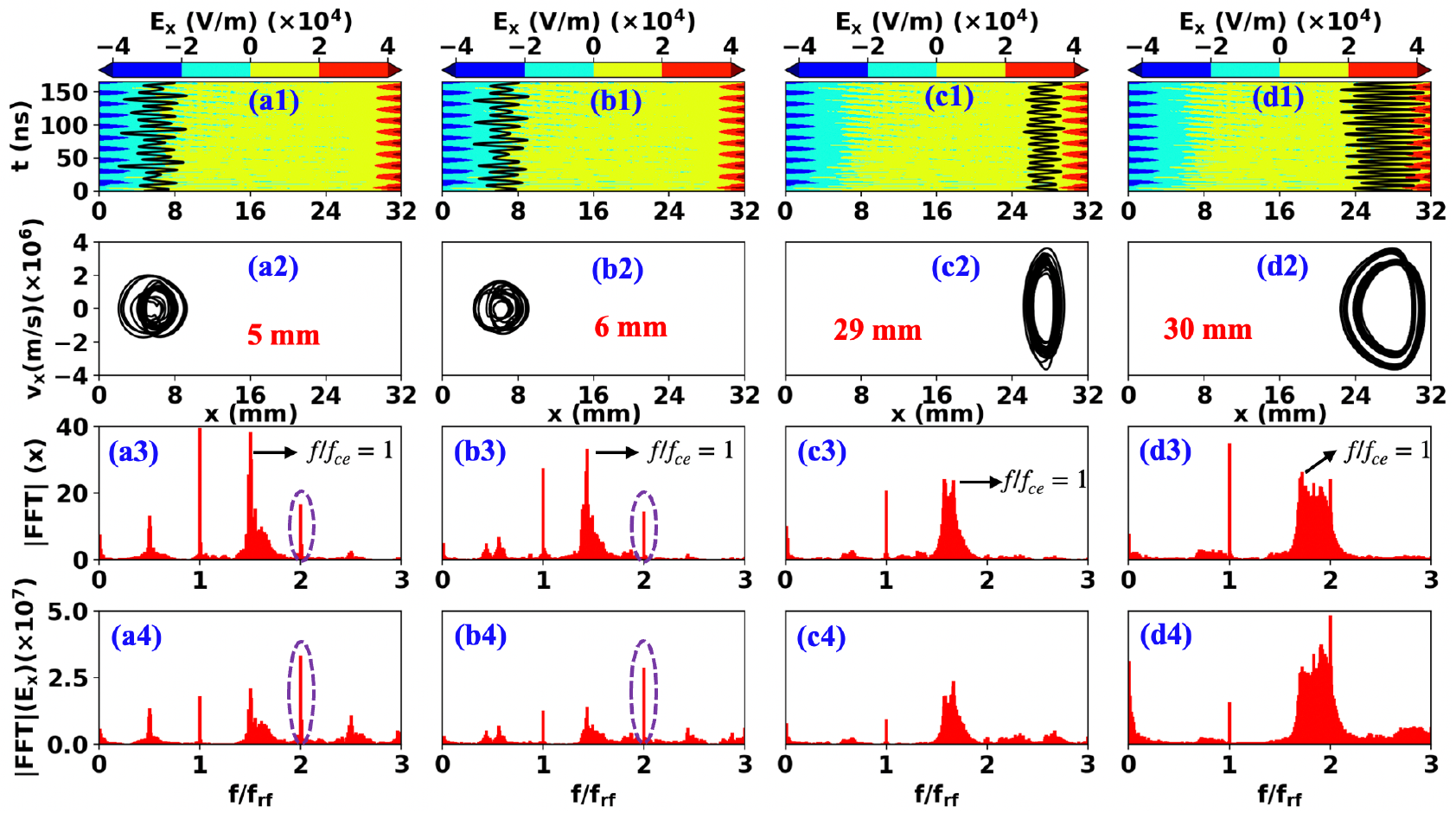}
    \caption{Analysis of test electron trajectories, having an initial energy of 2 eV, introduced at $x= 5$ mm (first column), $x= 6$ mm (second column), $x= 29$ mm (third column) and $x= 30$ mm (forth column) for $\mathrm{r = 2.5}$ ($\mathrm{26.7 \ Gauss}$). The top row (a1-d1) shows particle trajectories with $\mathrm{E_x(x,t)}$ and the second row (a2-d2) shows $\mathrm{x}$–$\mathrm{v_x}$ phase space. FFT spectrum of $x-t$ timeseries is shown in third row (a3-d3) whereas the forth row (a4-d4) shows FFT of electric field seen by the test particle.}  
    \label{trajectory}
\end{figure*}
In this subsection, we analyze test particle trajectories to further investigate the origin of EBW in these discharges. An equivalent explanation for the excitation of EBW driven by the displacement current, caused by sheath oscillations, is given in the companion paper submitted to Physical Review Letters \cite{gautam2025resonantly}. We have seen that the fundamental ($n = 1$) mode of EBW corresponds to the second harmonic of the RF frequency. Therefore, in order to identify the region where the EBWs are excited, we focus mainly on the signatures of this frequency in the dynamics of the test particles. {To investigate this further, we analyze the trajectories of test electrons with an initial energy of 2 eV, positioned near the sheath edges of both the powered and grounded electrodes. The analysis is carried out for $r = 2.5$, a condition under which the second harmonic generation is the most pronounced.}\\
The results of the trajectory analysis of the test particles are shown in Fig.~\ref{trajectory}. The columns of this figure correspond to the trajectories of four test electrons, of which two are initially positioned near the powered sheath ($x = 5, 6$ mm) and the remaining two near the grounded sheath ($x = 29, 30$ mm). The top row (a1-d1) shows the spatio-temporal profile of the electric field for 10 RF cycles along with the test particle trajectories. Each trajectory, shown in black line, is essentially a `time-series data', on which FFT analysis can be performed to identify the presence of second harmonics. The trajectories in the phase-space are shown in the second row (a2-d2) of the figure. The Fourier spectrum of $\mathrm{x-t}$ time series is plotted in the third row (a3-d3). Finally, in the fourth row (a4-d4), the Fourier spectrum of time series of an instantaneous electric field seen by the test particle is shown. These test particle trajectories are computed up to 100 RF cycles after the steady-state is achieved. Note that in Fig.~\ref{trajectory} (a1-d1, a2-d2), we have plotted the trajectories for 10 RF cycles for better visualization, but the FFT analysis (a3-d3, a4-d4) is carried out with trajectory data of 100 RF cycles. \\
The motion of test electrons near the magnetized sheath is intrinsically complex as a result of its interactions with the electric field of the oscillating sheath. These gyrating electrons experience highly non-linear time varying electric fields involving fundamental frequency ($f_{rf}$) and its harmonics. The phase space trajectories, shown in the second row, demonstrate this behavior. In particular, electrons near the powered sheath (a2,b2) undergo more chaotic motion compared to the grounded sheath (c2,d2). This explains the stronger ionization near the powered electrode side (shown in Fig.~\ref{ne_vol1}). \\
Because test electrons respond to the electric fields present in the sheath region, FFT analysis of their position/velocity time series can be useful in identifying the characteristic temporal scales associated with sheath dynamics \cite{markidis2020automatic}. In particular, the presence of distinct peaks in both test particle response and local electric field spectra indicates coherent interaction with electric fields. However, the broadband features signify stochastic interactions. The broadband spectra, seen in FFT of the test particle trajectories shown in the third row (a3-d3), correspond to such interactions within the sheath. Note the presence of cyclotron frequency ($f_{ce}$) within these broadband spectra, indicating the stochastic motion of magnetized electrons in the presence of oscillating sheath fields. In general, we find that the broadband becomes narrower as we move away from the edge of the sheath due to weakening of sheath electric fields. This trend is clearly visible on the grounded electrode when we compare Fig.~\ref{trajectory}(c3) (x = 29 mm) with Fig.~\ref{trajectory}(d3) (x = 30 mm).\\
Finally, the comparison of the third (a3-d3) and fourth (a4-d4) rows of Fig.~\ref{trajectory} shows a strong correlation between periodicity in the motion of the test electron and the harmonics of the electric field sampled by the same electron. For example, we see a distinct peak at fundamental RF frequency in both position and electric field spectra of these test electrons, indicating the ubiquitous presence of applied frequency in the vicinity of both the sheath. Interestingly, a distinct peak in the second harmonic (encircled by dashed line), corresponding to the EBWs frequency, is predominantly seen only near the powered sheath region.  Note that the position and electric field FFT of the electron, initially placed at the outer edge of the grounded sheath (c3,c4 corresponding to $x = 29$ mm), do not show the presence of a second harmonic of the applied RF frequency. This trend continues as we move further away from the sheath near the grounded electrode. In contrast, the presence of the second harmonic in both spectra is consistently found in the region near the powered sheath. This strongly suggests excitation of EBWs near the powered sheath region. These waves propagate further towards the grounded electrode. \\
It should be noted that the propagation of EBWs from powered to grounded electrode seems to be in conflict with spatio-temporal electric fields shown in Fig.~\ref{electric_field}. In other words, the tilt in the electric field spatio-temporal plots shown in Fig.~\ref{electric_field} suggests the waves are excited near the grounded sheath and propagate towards the powered sheath, which is in contrast with the conclusions drawn from test particle analysis. This paradox can be resolved when we consider the relation between the group velocity and the phase velocity of EBWs. In the parameter regime studied in this paper, the EBWs behave like backward propagating waves \cite{stix1992waves}, characterized by opposite directions of phase and group velocities. Therefore, propagation of the constant phase front in Fig.~\ref{electric_field} appears from grounded to powered electrode. On the other hand, energy coupling through the group velocity occurs from the powered electrode to the grounded electrode. This is consistent with Fig.~\ref{e_heating}, where we see time-averaged electron heating diminishing from powered to grounded electrode in the presence of EBWs. \\
Finally, we comment on the role of density gradient in the excitation of EBWs. Our results show a stronger excitation of EBWs in the presence of a density gradient in the bulk plasma. This trend is consistent with the correlation between density scale-height and EBW coupling reported in theoretical and experimental studies in magnetic fusion physics \cite{mjolhus1984coupling,peter2007electron, kohn2011full}. In particular, these studies have demonstrated that the coupling efficiency of electromagnetic to electrostatic waves is expressed in terms of the Budden parameter \cite{budden1988propagation}, which is expressed in terms of the density scale-height. In the case of CCP discharges, it can be assumed that similar considerations will hold when the second harmonic of the applied voltage is converted to the EBW.  \\
\section{4. \ CONCLUSION}
Previous studies have demonstrated that magnetized CCP discharges can effectively enhance plasma density, particularly near the condition $2\,\mathrm{f_{ce} = f_{rf}}$~\cite{patil2022electron}. In this work, we have investigated the origin of asymmetry in plasma density and ionization across the range $\mathrm{r = 2.0}$ to $\mathrm{r = 3.5}$. Our results show that the ion flux is maximized at the grounded electrode, while the electron flux peaks at the powered electrode. At $\mathrm{r = 2.0}$ ($\mathrm{f_{ce} = f_{rf}}$), electrons interact with the sheath in phase, minimizing losses at both electrodes and resulting in symmetric density and ionization. In this regime, the powered sheath does not collapse completely.\\
In contrast, during asymmetric discharges, electron-sheath interactions at the powered side generate energetic electrons. Some are lost to the electrode, while others reflect and traverse the bulk. Many of these electrons accumulate near the grounded sheath as a result of insufficient energy to overcome its potential, causing a localized density peak. Meanwhile, ions—due to their inertia—respond to the time-averaged bulk electric field and are directed toward the grounded electrode. This leads to electron and ion density peaking near the grounded side, while ionization remains highest near the powered sheath due to localized electron heating. As a result, the time-averaged electron flux is greater at the powered electrode, while the ion flux is higher at the grounded side.\\
Finally, in this regime, where the discharge becomes asymmetric, we observe the excitation and propagation of Bernstein waves. Specifically, the second harmonic of the RF frequency resonantly excites the fundamental mode of the Bernstein wave. The presence of a second harmonic in the spectral component of the test particle trajectory confirms this hypothesis. These waves travel inside the overdense bulk plasma, as is expected from the electrostatic nature of the Bernstein waves. The existence of these waves opens up exciting possibilities for heating the bulk plasma using waves. This will be especially useful in the low density regime, where ohmic heating of bulk plasma is very weak.  

Acknowledgments: IDK research was supported by the US Department of Energy, Office of Fusion Energy Science, under contract $\#$ DE-AC02-09CH11466 as a part of the Princeton Collaborative Low Temperature Plasma Research Facility (PCRF). The authors also thank the NEUMANN and ANTYA HPC facilities at CEBS (Mumbai) and IPR (Gandhinagar), respectively. The authors would also like to thank the anonymous reviewer for insightful constructive criticism that led to significant improvements in the manuscript.

\bibliographystyle{apsrev4-2}
\bibliography{apssamp} 

\end{document}